\documentclass[sigplan,10pt,nonacm,balance=false]{acmart}

\setcopyright{none}
\renewcommand\footnotetextcopyrightpermission[1]{}

\usepackage{tabularx}
\usepackage{graphicx}
\usepackage{multirow}
\usepackage{enumitem}
\usepackage{booktabs}
\usepackage{amsmath}
\usepackage[normalem]{ulem}
\usepackage{mathtools}
\usepackage{xspace}
\usepackage{makecell}
\usepackage{xcolor}
\usepackage{colortbl}
\usepackage{algorithm}
\usepackage{algorithmic}
\usepackage[capitalize,noabbrev]{cleveref}

\newcommand{\sysname}{ROSS\xspace}
\newcommand{\sysnames}{ROSS's\xspace}

\newcommand{\ycao}[1]{}
\newcommand{\bohan}[1]{}
\definecolor{mypurple}{RGB}{102,0,153}
\newcommand{\yongchao}[1]{}

\definecolor{ycnew}{RGB}{205,85,85}
\definecolor{ycold}{RGB}{145,105,105}
\newcommand{\ycadd}[1]{#1}
\newcommand{\ycdel}[1]{}
\newcommand{\ycrep}[2]{#2}

\newcommand{\para}[1]{\par\noindent\textbf{#1}}

\usepackage[disable]{todonotes}

\begin{document}

\title{ROSS: Dual-Plane Simulation for Efficient and Accurate LLM Serving Systems}

\author{Ye Cao}
\affiliation{%
  \institution{Fudan University}
  \country{}}
\email{ycao24@m.fudan.edu.cn}

\author{Bohan Zhao}
\affiliation{%
  \institution{Unaffiliated}
  \country{}}
\email{zhaobh23@tsinghua.org.cn}

\author{Yongchao He$\textsuperscript{*}$}
\affiliation{%
  \institution{Unaffiliated}
  \country{}}
\email{yongchao-he@outlook.com}

\makeatletter
\def\@fnsymbol#1{\ensuremath{\ifcase#1\or *\or \dagger\or \ddagger\or
   \mathsection\or \mathparagraph\or \|\or **\or \dagger\dagger
   \or \ddagger\ddagger \else\@ctrerr\fi}}
\makeatother

\thanks{$^*$Corresponding author.}

\begin{abstract}
Modern large language model (LLM) serving systems expose a large and evolving design space spanning parallelism strategies, scheduling policies, and resource budgets, making system-level simulation a cost-effective tool for evaluating these choices.
Existing simulators rely on either simulator-side abstractions or full-runtime replay, sacrificing either fidelity or efficiency.
We trace this to a \emph{computation--control asymmetry}: model execution is compute-intensive yet predictable, whereas serving control is lightweight but highly state-dependent.
ROSS decouples these concerns into two planes.
A \emph{data plane} predicts per-iteration latency via a stage-wise regression model that captures the resolved effects of asynchronous CPU--GPU overlap without reconstructing fine-grained dependencies.
A \emph{control plane} invokes the serving framework's load-balancing, scheduling, and memory-management modules over simulated state without simulator-side re-implementation or full-runtime replay.
Evaluated across NVIDIA H200/B200, seven models (up to 671B parameters) and two serving frameworks (vLLM, SGLang), ROSS achieves below 6\% median error for workload end-to-end duration and time per output token (TPOT), and over $11\times$ single-instance speedup over real execution.
In two case studies, ROSS conducts a 228-candidate configuration search in 70 minutes on a CPU server, avoiding 262.4 B200 GPU-hours of exhaustive hardware evaluation, and identifies an SGLang post-generation bottleneck whose fix reduces workload end-to-end duration by 35.8\% at high concurrency.
\yongchao{[abstract precision] Two clauses overstate or obscure the evidence. The stage-wise model does not explicitly reconstruct CPU--GPU dependencies; it learns boundary-aligned latencies containing their resolved effects. The control-fidelity clause is also not interpretable without a metric: state the measured agreement (currently the same 1,280 non-empty iterations plus the reported trajectory comparison) and name the mismatched baseline, or add a quantitative trajectory-distance measure before making a stronger claim.}
\end{abstract}

\maketitle
\makeatletter

\fancypagestyle{standardpagestyle}{
  \fancyhf{}
  \renewcommand{\headrulewidth}{0pt}
  \renewcommand{\footrulewidth}{0pt}

  \fancyhead[LO,LE]{\ACM@linecountL}
  \fancyhead[RO,RE]{\ACM@linecountR}

  \fancyfoot[C]{\footnotesize\thepage}
}

\pagestyle{standardpagestyle}
\makeatother

\yongchao{[global citation and evidence audit] Please perform a paper-wide claim-to-evidence pass. Several statements about external framework behavior, production practices, hardware compatibility, policy evolution, simulator availability, and cloud pricing are either uncited or supported only by a nearby citation that does not establish the specific claim. Cite versioned documentation, source code, release notes or commits, artifact repositories, or workload studies as appropriate; include the access date, region, and pricing model for time-sensitive cost claims. Conversely, label paper-generated observations explicitly as `in our evaluation,' and support causal interpretations with direct ablations or narrower wording rather than adding unrelated citations.}

\yongchao{[global visual consistency] The figures and tables do not yet read as one coherent visual system. Establish and apply a publication-wide specification for single- and double-column dimensions, minimum embedded font size, line weight, color semantics, terminology, metric labels, and caption structure. Rework low-information or undersized figures, make colors and system identities consistent across panels, annotate small heatmaps numerically, and keep each float adjacent to the claim it supports rather than forcing placement with `[H].' Also rebalance pages that stack several dense floats or leave large unused regions, and standardize table font size, alignment, abbreviations, and standalone captions.}

\section{Introduction} \label{sec:intro}

Modern large language model (LLM) serving systems~\cite{yu2022orca,kwon2023vllm,zheng2024sglang,nvidia_tensorrt,aminabadi2022deepspeed,zhong2024distserve} operate in \ycrep{a complex, high-dimensional design space: model architecture, hardware topology, parallelism strategy, scheduling policy, and memory budget jointly determine serving performance}{a large and rapidly evolving design space spanning hardware topology, parallelism, scheduling, and memory management}~\cite{miao2026serving_survey,zhou2024llm_inference_survey}.
These choices interact across computation, communication, and memory, shifting bottlenecks \ycrep{as LLM serving systems evolve. The throughput gap between a poorly matched deployment and an optimal one can exceed $1.9\times$~\cite{bambhaniya2025endtoend}, making design-space exploration a prerequisite for meeting latency and throughput Service-Level Agreements (SLAs) rather than a secondary optimization. However, evaluating this space on real hardware is prohibitively expensive: even an 8-GPU deployment can expose 200+ }{across serving designs. Serving-system designers and operators therefore need efficient design-space exploration as workloads and serving stacks evolve, often under service-level agreement (SLA) constraints~\cite{patel2024splitwise,splitwisesim}. Yet evaluating these alternatives directly on hardware is costly: each design requires deployment, warm-up, workload replay, and metric collection~\cite{agrawal2024vidur,fu2024serverlessllm,zeng2025medusa}; benchmarking 358 Qwen3-32B~\cite{qwen2025qwen3} configurations takes 35.4 hours on an 8$\times$H100 SXM node~\cite{aiconfigurator2026}.} \yongchao{[application scenarios] The target use case is still abstract. Please describe one concrete operator workflow: who uses ROSS, at what point in the serving lifecycle, which inputs are available, which decision it produces, and how often that decision is revisited. For example, is the primary scenario offline pre-deployment selection of parallelism, memory, and PD topology for a long-running service, periodic capacity replanning as traffic changes, or framework debugging? Naming the primary scenario will clarify which fidelity and speed requirements actually matter.}
\emph{System-level simulation} offers a faster and cheaper way to predict the performance of untested designs and support what-if analysis~\cite{bambhaniya2025endtoend} \ycrep{for deployment planning, capacity planning and performance diagnosis without risking production regressions.}{without exercising every design on physical hardware.}
\yongchao{[search motivation] The argument jumps from `there are 200+ candidates' to the need for broad search, but an operator could cheaply prune most configurations and accept a near-optimal result. Explain when exhaustive or near-exhaustive exploration is necessary, and compare it with a pruning or heuristic baseline in terms of decision regret. The long service lifetime is the missing economic argument: even a 1\% throughput or efficiency gap can accumulate into substantial GPU-hours, capacity, and energy at fleet scale, so quantify the break-even horizon at which finding the better configuration repays the additional search effort.} \ycdel{candidates, and each evaluation requires warm-up, workload replay, and tens of minutes of metric collection~\cite{agrawal2026revati}.} 
\yongchao{[why-now] The opening establishes that design-space exploration is expensive, but never answers `why now?' Add one evidence-backed sentence explaining which recent serving trends---for example PD disaggregation, MoE routing, or a faster cadence of framework policies---expand the configuration space or increase control sensitivity beyond what prior simulators were designed for.}

Delivering simulators that are simultaneously \emph{efficient} and \emph{high-fidelity} remains an open problem. Fast design-space exploration favors low-overhead performance estimation so that \ycdel{thousands of }candidate configurations can be screened quickly~\cite{agrawal2024vidur,aiconfigurator2026,liu2025llmshare}. Validating deployment decisions, in contrast, demands accurate modeling of request-level scheduling, \ycrep{memory fragmentation, }{key--value (KV) cache }and resource contention~\cite{sun2024llumnix}. This tension yields a speed--accuracy trade-off~\cite{casanova2014simgrid}: \ycrep{higher fidelity expands the simulated state space and event volume, reducing scalability}{ preserving more execution and runtime details generally increases simulation cost}.

Existing simulators navigate this trade-off by deciding how much of the serving stack to model and to retain. Approaches that favor speed rely on coarse latency models and simulator-side control logic, enabling scalable exploration but risking divergence from current framework behavior.
Approaches that prioritize fidelity either retain most of the native serving stack, preserving framework behavior but inheriting its runtime footprint~\cite{da2026llmemu,agrawal2026revati}, or reconstruct system behavior through detailed profiling and re-implemented framework logic, incurring substantial profiling and \ycrep{adaptation costs}{simulator-side adaptation}~\cite{agrawal2024vidur,cho2024llmservingsim,bambhaniya2025endtoend,zhang2024llmcompass}. \yongchao{[orphan claim] This sentence asserts a field-wide diagnosis (`the lack of a principled simulation boundary') without evidence. Is this your observation or an established finding? Either cite prior work that reaches the same conclusion, or rephrase as `We observe that...' to signal this is your framing.} \ycrep{A key limitation underlying both directions is the lack of a principled simulation boundary between modeling execution dynamics and preserving framework behavior. Existing }{Taken together, these designs expose a common limitation in dealing with execution modeling and framework-behavior preservation.
Different modeling levels balance execution fidelity against profiling and simulation cost.
Similarly, for framework behavior, simulator-side approximations of state-sensitive control logic can cause resource-footprint divergence, while full-stack execution preserves native behavior at the cost of runtime overhead.} Table~\ref{tab:comparison} summarizes these architectural trade-offs \ycadd{among representative LLM serving simulators}.

\begin{table}[t]
\centering
\caption{\ycadd{Comparison of LLM serving simulators in terms of how iteration latency is predicted (Execution), scheduling and memory-management decisions are obtained (Control), and request traffic is modeled (Traffic).}}
\vspace{-1em}
\label{tab:comparison}
\small
\begin{tabular*}{\columnwidth}{@{\extracolsep{\fill}}lccc@{}}
\toprule
& Execution & Control & Traffic \\
\midrule

\makecell[c]{Vidur~\cite{agrawal2024vidur}}
& \shortstack[c]{Op.-level}
& \shortstack[c]{Re-impl.}
& \shortstack[c]{Open-loop} \\

\makecell[c]{Revati~\cite{agrawal2026revati}}
&
\makecell[c]{Pluggable; \\ default Vidur's}
&
\makecell[c]{Native\\(Time-warp)}
&
\makecell[c]{Open-loop \\ eval.}
\\

\makecell[c]{AIConf.~\cite{aiconfigurator2026}}
&
\makecell[c]{Profiled \\ op.-level}
&
\makecell[c]{Analytical approx.}
&
\makecell[c]{Static}
\\

\makecell[c]{\textbf{\sysname}}
&
\makecell[c]{\textbf{Stage-wise}\\\textbf{(learned)}}
&
\makecell[c]{\textbf{Native}\\\textbf{modules}}
&
\makecell[c]{\textbf{Closed-loop}}
\\

\bottomrule
\end{tabular*}
\end{table}

\yongchao{[skim test] A skim reader sees this table before reading any design section. The column headers (`Execution', `Control', `Traffic', `Scaling') are too terse to convey what each dimension means or why it matters. Add a 2-line caption explaining: e.g., `Execution: how iteration latency is predicted. Control: how scheduling/memory decisions are obtained. Traffic: whether the simulator handles online/closed-loop arrivals. Scaling: how runtime grows with design-space size.' Also, the one-word cell values (`Operator', `Analytical', `Stage-wise') are jargon that only makes sense after reading the paper. Consider short glosses, e.g., `Stage-wise (learned)'.}

\para{Key insight.}
We observe that \emph{execution} and \emph{control} contribute to serving performance in fundamentally asymmetric ways. Model execution, namely the GPU-side forward pass, is computationally intensive yet largely \emph{predictable} given batch shape, model architecture, and hardware characteristics, as its latency follows regular compute- and memory-bound patterns~\cite{williams2009roofline,lee2025forecasting,agrawal2024vidur}. Together with CPU-side preparation and completion costs, execution primarily determines how much time elapses during each serving step. Control logic, including scheduling, batching, KV-cache allocation, and preemption, is computationally lightweight but highly state-sensitive: its decisions determine how request and resource state evolve, and differences in these decisions can propagate to subsequent iterations~\cite{yu2025ic,kwon2023vllm,sun2024llumnix}.
This asymmetry suggests a principled simulation boundary. \ycrep{The execution process}{Execution timing} can be modeled predictively, whereas state-sensitive control behavior should be obtained directly from the framework's native modules. Crucially, obtaining native decisions requires neither simulator-side control re-implementation nor full-stack runtime replay.

\para{Approach.}
We operationalize this boundary by decoupling how execution and control are simulated. \sysname\footnote{\sysnames code is available at \url{https://anonymous.4open.science/r/ross-34CF}} learns execution dynamics from \ycrep{sparse traces}{real-hardware serving traces} while invoking framework-native decision logic through narrow interfaces.
On the \emph{data plane} (execution modeling), \sysname \ycrep{decomposes each iteration into pre-forward, forward, and post-forward stages that explicitly absorb CPU--GPU synchronization gaps, and trains a lightweight regressor (XGBoost~\cite{chen2016xgboost}), avoiding kernel-by-kernel execution modeling.}{
constructs pre-forward, forward, and post-forward timing targets around model execution that capture the resolved effects of asynchronous CPU--GPU overlap, and trains XGBoost regressors~\cite{chen2016xgboost} on these targets, avoiding fine-grained execution reconstruction.
}
On the \emph{control plane} (scheduling and resource management), \sysname invokes the framework's load-balancing, scheduling, and memory-management implementations (collectively, \emph{native control modules}) through three narrow interfaces, without simulator-side re-implementation or full-stack replay of the framework runtime.
\ycrep{
A virtual hardware abstraction layer (vHAL) bridges the two planes: it intercepts execution calls and returns predicted latencies in $O(1)$ time, advancing a virtual clock without physical devices.
}{
Within the control plane, a virtual hardware abstraction layer (vHAL) replaces physical model execution with predicted completion events on the simulator's virtual timeline.
}
\para{Results.}
We evaluate \sysname across NVIDIA H200 and B200 platforms, \ycrep{seven}{ten} models in total (8B--671B, dense and mixture-of-experts (MoE)\ycadd{; seven in the main
evaluation and three in held-out evaluations}), two \ycrep{production}{serving} frameworks (vLLM~\cite{kwon2023vllm}, SGLang~\cite{zheng2024sglang}), and both co-located and prefill--decode (PD) disaggregated modes.
\sysname achieves below 6\% median \ycrep{percentage error (PE)}{error} for workload end-to-end (E2E) duration and time per output token (TPOT), 8.3\% median error in mean time to first token (TTFT) on long-context prefill workloads (2.5K--20K-token prompts), and over $11\times$ simulation speedup over real execution\ycdel{; it also follows a modified vLLM preemption policy's KV-cache footprint while mismatched control paths diverge}.
Two case studies demonstrate operational utility beyond prediction accuracy: \ycrep{ A 228-candidate Pareto search completes in 70 minutes on a CPU-only server at 1,258$\times$ lower cost (\$1.95 vs. \$2,457 on AWS) than exhaustive hardware evaluation}{ A 228-candidate Pareto search completes in 70 minutes on a CPU-only server compared with 262.4 B200 GPU-hours for exhaustive hardware evaluation\ycdel{ (1,258$\times$ lower cost, \$1.95 vs. \$2,457 on AWS)}}, while a \ycrep{stage-level diagnosis identifies a SGLang bottleneck}{simulation-guided diagnosis reveals an SGLang post-generation bottleneck} whose fix reduces E2E duration by 35.8\% under high-concurrency workloads.
\yongchao{[cost metric] Dollar cost should not be the primary headline because cloud prices vary by provider, region, discount model, and time. Lead with stable resource-normalized quantities---GPU-node-hours or GPU-hours avoided, CPU core-hours consumed, wall-clock search time, and number of real-hardware trials---and retain the AWS estimate only as a dated secondary illustration with its pricing assumptions.}



\section{Background and Related Work}
This section summarizes the iteration-level execution model of modern serving frameworks (\S\ref{sec:2_1}) and reviews existing simulation approaches (\S\ref{sec:2_2}). In this paper, \emph{system-level serving simulation} denotes reproducing a serving framework's request- and resource-state evolution over virtual time, rather than merely predicting execution dynamics.

\subsection{Iteration-level Execution in LLM Serving} \label{sec:2_1}

Modern LLM serving frameworks drive inference in an iteration-based loop~\cite{yu2022orca,agrawal2024sarathi}: the engine selects a runnable set of active requests and \ycrep{advances each by}{schedules each for} one execution step (one decode token or a chunk of prompt prefill), making the iteration the basic unit for scheduling and accounting.

Within one iteration, the observed wall-clock time \ycrep{emerges from a sequence of interleaved \emph{sub-steps} that alternate between the CPU and the GPU:}{reflects several interleaved \emph{functional components} spanning CPU, GPU, and network activity:}
\begin{enumerate}[leftmargin=*, label=\arabic*., itemsep=0pt, topsep=0pt, parsep=0pt]
\item \textit{Preparation} (CPU): The scheduler selects which requests to run, allocates KV-cache memory blocks, and assembles metadata for the next batch.
\item \textit{Execution} (GPU): The batched forward pass---attention~\cite{dao2022flashattention} and MLP kernels---runs on the device.
\item \textit{Completion} (CPU): Results are synchronized back to the host; the scheduler updates request state (token append, termination check, block release).
\item \textit{Transfer} (network, optional): In prefill--decode disaggregated deployments, KV-cache blocks are transferred from the prefill worker to the decode worker before \ycadd{decode} execution can proceed~\cite{zhong2024distserve,qin2026mooncake}.
\end{enumerate}
\yongchao{[local logic] The next paragraph correctly says these activities overlap, but the numbered list still implies an order in which transfer follows completion. In a PD path, transfer is a readiness condition for decode execution; label the list as functional rather than temporal, or show separate co-located and disaggregated timelines.}
Crucially, these sub-steps are not executed as a strict sequence. Modern engines proactively overlap the CPU-side preparation \ycrep{of the \emph{next} iteration with the GPU execution of the \emph{current} one}{for a later serving iteration with GPU execution of an earlier batch~\cite{zhu2025nanoflow}}, so iteration latency is governed by a dependency-driven critical path rather than a linear sum of component costs. \ycrep{This coupling motivates more principled modeling than monolithic-step or linear-aggregation abstractions can provide.}{This coupling therefore limits the accuracy of monolithic-step and linear-aggregation abstractions as serving conditions change.}

\ycadd{\para{Control-state loop.}
The scheduling, allocation, and completion steps above operate on a persistent server-side state that includes request queues, request progress, KV/block allocation, and per-worker load. Scheduling decisions read this state to form the next batch, while completion updates write back request and resource changes that become the input state for subsequent iterations. Under concurrency-limited traffic, request completion can also release a client-side semaphore slot and thereby enable a later admission, coupling the server control loop with request generation~\cite{sglang_bench_serving}.}

\subsection{Related Work} \label{sec:2_2}

Prior work differs in how it approximates model execution and serving control flow.

\para{Estimating model execution.}
\textbf{Kernel-level}~\cite{peng2025reallm,yao2026kernelsight}
and \textbf{operator-level}~\cite{cho2025llmservingsim2,feng2025frontier,aiconfigurator2026}
approaches decompose execution into fine-grained component costs. They capture detailed hardware effects, but profiling and calibration requirements vary by approach; recent work reduces measurement costs through profile reuse~\cite{kim2026dooly} and cross-generation prediction~\cite{yao2026kernelsight}.
Coarser \textbf{token-/iteration-level} models~\cite{lin2024apex,wu2025tokensim,liu2025llmshare} can reduce profiling cost, but monolithic prediction may miss intra-iteration dynamics.
\ycrep{Pipelining (e.g., CUDA Graphs~\cite{nvidia2026cudagraphs})}{Asynchronous pipelining} further weakens linear-aggregation assumptions.
\yongchao{[misplaced evidence] These H200 error numbers appear to be produced by our §5 evaluation, but here they read as results reported by Vidur. Move them to §5.2 and keep Related Work methodological, or write `In our evaluation (§5.2)...' so evidence ownership is unambiguous.}
\yongchao{[technical precision] CUDA Graphs primarily capture and replay launch graphs; they are not themselves an example of CPU--GPU pipelining. Please name the actual asynchronous overlap mechanism here, or explain precisely how graph replay changes the linear-aggregation assumption.}

\para{Emulating serving control flow.}
\textbf{Static} approaches use \ycrep{closed-form}{analytical} models~\cite{aiconfigurator2026}; \textbf{trace-driven} methods replay historical executions~\cite{cho2025llmservingsim2, peng2025reallm}; \textbf{event-driven} simulators model state transitions over time~\cite{banks2010des,agrawal2024vidur, feng2025frontier}.
Their fidelity depends on how accurately they represent state-dependent serving control: simplified heuristics miss SLA-critical behaviors (e.g., preemption triggers~\cite{kim2024effect,sheng2024fairness}, chunked prefill~\cite{agrawal2025efficient}, KV-cache compression~\cite{liu2024cachegen}), while faithful re-implementation incurs high maintenance cost as frameworks evolve.
\ycrep{\textbf{Time-warp emulation} takes a different approach: it runs the \emph{entire} real framework stack and skips only GPU kernel execution via virtual time advancement.}{\textbf{Native-runtime emulation} retains the \emph{entire} serving runtime while substituting GPU execution. LLM-Emu~\cite{da2026llmemu} uses profile-sampled forward delays in wall-clock time, whereas Revati's \emph{Time-warp emulator}~\cite{agrawal2026revati} skips GPU kernel execution through coordinated virtual-time advancement.}
\ycrep{Both avoid re-implementation drift but inherits the \ycdel{full} CPU overhead of the production code path (Python interpreter, multi-process IPC, CUDA driver stubs), limiting simulation throughput for large-scale design-space sweeps.}{Both avoid re-implementation drift but retain production CPU work, limiting simulation throughput for large-scale design-space sweeps.}
\ycrep{
Moreover, time-warp emulation relies on a global Timekeeper that requires every active component to pre-announce its next event timestamp, where a client cannot determine its next arrival time until the server responds. This can block progress or introduce additional coordination under concurrency-constrained traffic.
}{Time-warp emulation relies on a global Timekeeper that coordinates virtual-time advancement across runtime processes. Under concurrency-constrained traffic, completion-dependent admissions cannot always be determined in advance, requiring additional coordination.
}

\ycrep{As summarized in Table~\ref{tab:comparison}}{Overall}, \sysname adopts a different simulation boundary from prior work: it models execution timing and invokes native decision and update rules on simulated state, without fine-grained execution reconstruction, simulator-side policy re-implementation, or full-runtime replay.

\section{Challenges} \label{sec:challenges}

This section analyzes the fundamental system properties that make LLM serving simulation difficult: \ycrep{efficiently modeling the non-linear execution dynamics}{accurately modeling execution timing under asynchronous overlap and limited internal observability} (\S\ref{sec:3_1}), and faithfully \ycrep{reproducing}{simulating} control flow under state-sensitive decisions, evolving policies, and closed-loop traffic (\S\ref{sec:3_2}).

\subsection{Obstacles to Accurate and Cost-Efficient Execution Modeling} \label{sec:3_1}

Accurately and cost-efficiently predicting the latency of a single iteration faces \ycdel{following} two interrelated \ycrep{tensions}{challenges}.

\para{Latency overlap under asynchronous execution.}
Asynchronous pipelining inherent in modern serving engines (\S\ref{sec:2_1}) induces \emph{latency overlap}: the observed iteration time cannot be obtained by directly summing isolated CPU- and GPU-side durations, but is determined by a dependency-driven critical path across preparation, execution, and synchronization.
\ycrep{
Capturing this behavior requires a simulator to represent cross-device dependencies (e.g., launch and synchronization boundaries) and to model when overlap is feasible versus when it is blocked by resource availability.
}
{Accurate prediction therefore requires capturing the resolved timing effect of cross-device dependencies (e.g., launch and synchronization boundaries) and resource availability.
}
\yongchao{[missing edge] Section~3 says an accurate simulator must represent cross-device dependencies, whereas Section~4 says \sysname deliberately avoids reconstructing them. Reframe the requirement as capturing the \emph{resolved effect} of those dependencies, then explain why boundary-level targets satisfy it; otherwise the challenge demands a mechanism \sysname does not provide.}
\ycrep{
Many simulators \yongchao{cite some papers} adopt a sequential abstraction and compute iteration latency by summing isolated components; this linearization ignores overlap and bottleneck shifts, leading to inaccurate iteration timing as overlap and bottlenecks shift across operating regimes.
}{Predicting isolated component costs is insufficient unless their composition preserves the overlap realized by the serving runtime. At the other extreme, predicting the entire iteration as a single target must absorb heterogeneous host, device, and communication effects into one mapping.
}

\para{Measurement blind spots.}
Execution performance modeling requires ground-truth measurements that map execution steps to latency. \ycrep{Yet, production optimizations---CUDA Graphs~\cite{nvidia2026cudagraphs}, FlashAttention~\cite{dao2022flashattention}, and kernel fusion---fuse operators and encapsulate execution in opaque monolithic contexts, effectively blinding host-side timers to internal structure.}{Yet production optimizations such as CUDA Graphs~\cite{nvidia2026cudagraphs}, fused kernels such
as FlashAttention~\cite{dao2022flashattention}, and operator fusion reduce the granularity visible
to host-side timing.} To penetrate this ``black box,'' practitioners often disable key optimizations (e.g., reverting to eager execution) or add heavy instrumentation (e.g., CUDA event injection)~\cite{wu2026stria}. Both options can perturb scheduling and kernel behavior, producing distorted measurements that fail to reflect the deployment environment and undermine simulation fidelity.
\ycadd{\sysname therefore places probes at existing framework-visible boundaries around model execution, without instrumenting the optimized device-side execution path (\S\ref{sec:4_2}).}
\yongchao{[challenge-to-design bridge] The proposed solution also instruments the framework, but the paper never closes this loop. Please explain why its boundary instrumentation preserves CUDA Graphs, fusion, and asynchronous execution, and report a probed-versus-unprobed overhead sanity check.}

\subsection{\ycrep{Overheads of}{Obstacles to} High-Fidelity Control \ycrep{Logic Emulation}{Simulation}} \label{sec:3_2}

\ycrep{Serving control plane is inherently \emph{exact-state-sensitive}: request latency and capacity limits depend on the precise sequence of admission, preemption, and eviction decisions, which in turn depend on the instantaneous state of fragmented block tables, request queues, and KV-cache metadata. This property creates three obstacles for simulation.}
{As described in \S2.1, serving control reads and updates request and resource state across iterations, while request completion can also feed back into future admission. These dependencies make the control plane highly state-sensitive and create three challenges for simulation.}
\yongchao{[background-to-challenge bridge] Section~3.2 assumes substantially more control-plane machinery than Background establishes. Section~2 names scheduling, KV allocation, and completion, but does not define the server-side control state or its loop connecting admission, batch formation, block allocation/eviction, preemption, execution completion, and block release; it also does not show how that server loop couples to the client-side concurrency/semaphore loop. Add one compact state-transition diagram or running iteration in Background, use the same terminology here and in §4, and then point each of the three obstacles to a specific state or feedback edge. Otherwise, a reviewer who is not deeply familiar with vLLM/SGLang cannot tell why a one-step mismatch compounds or how closed-loop traffic interacts with, rather than simply belonging to, server control.}

\para{Resource-footprint divergence under approximation.}
Approximating control logic with simplified heuristics or separate re-implementations can introduce small per-step deviations in batch composition and memory allocation.
These deviations can propagate to subsequent iterations: a slightly different batch at iteration $i$ changes the KV-cache occupancy seen by the scheduler at iteration $i{+}1$, altering subsequent preemption and admission decisions.
Over multiple iterations, this compounding effect can therefore propagate through the serving process. \S\ref{sec:control_fidelity} demonstrates this effect \ycrep{: after modifying vLLM's preemption policy, native control reuse follows the resulting KV-cache footprint, whereas mismatched control paths diverge.}{ under a modified vLLM preemption policy, where retaining first-come, first-served (FCFS) changes 1,250 of 2,020 scheduler iterations relative to the modified-vLLM reference.} Reproducing control-plane behavior therefore requires keeping simulated decisions consistent with the framework's state-dependent logic.

\para{Rapid policy evolution.}
\ycrep{Serving control policies are an active and fast-moving research frontier, with frequent changes in batching, eviction, preemption, and scheduling strategies~\cite{kim2024effect,agrawal2025efficient,yu2025ic}.}{
Appendix Table~\ref{tab:vllm_evolution} shows five representative control-relevant changes between vLLM v0.10.0 and v0.11.2 over a four-month period, affecting scheduler ordering, KV interfaces, preemption state, and cache eviction.}

\yongchao{[evidence for evolution overhead] `Active and fast-moving' and the resulting maintenance burden are currently assertions supported by three topic citations, not evidence about change frequency or adaptation cost. Add a compact table over a defined time window that separates deployed framework changes from proposed research policies, with columns for date/version, release-note/PR/commit or paper identifier, changed control component, state or interface affected, and the simulator logic that a separate implementation would have to update. If available, include measured code churn or engineer effort for adapting one prior simulator; otherwise report auditable proxies such as the number of relevant upstream changes and touched modules. This would substantiate both the pace claim and the specific maintenance work that native-module reuse avoids.}
\ycrep{Any}{A} simulator that maintains a separate re-implementation of this logic faces a maintenance burden: each upstream framework update risks introducing behavioral drift that silently degrades simulation fidelity. \ycrep{The simulator must therefore \emph{reuse} native framework modules rather than duplicate them.}{This motivates reusing native framework modules rather than duplicating them.}

\para{Closed-loop traffic and concurrency feedback.}
Production LLM traffic is shaped not only by arrival rate and burstiness~\cite{wang2025burstgpt} but also by \emph{concurrency constraints}: Concurrency-limited clients can cap the number of in-flight requests, creating a feedback loop in which the client's next request time depends on the server's completion of a prior request.
Unlike open-loop traffic, where arrival times can be determined in advance, concurrency-constrained traffic couples future request admission to the completion of in-flight requests.
\ycrep{
A simulator that models concurrency-constrained traffic must therefore maintain a virtual concurrency pool that interacts with the scheduling loop on a \emph{virtual} timeline, without requiring physical blocking or global clock synchronization that would stall or degrade the simulation.
}{
A simulator must therefore evolve request admission and completion on a common virtual timeline so that future arrivals respond causally to simulated completions.
}

\subsection{\ycrep{Motivation}{Design Implications}}
\yongchao{[evidence-to-design ending] If the preceding analysis is revised to provide repeated-measurement evidence, direct falsification of prior abstraction classes, and a concrete policy-evolution survey, this ending should become much shorter. Do not re-argue the execution--control trade-off for a fourth time. Close Section~3 with a checkable evidence-to-requirement mapping---e.g., measured boundary-level predictability implies learned stage-wise execution; measured trajectory sensitivity implies native control reuse; observed closed-loop dependencies imply virtual-time traffic coordination---and then state in one sentence that \sysname must realize this dual-plane boundary. Section~4 can explain how. This makes the design feel inevitable from the data rather than rhetorically asserted.}
These challenges suggest two design implications: execution timing should be modeled at observable boundaries that retain the resolved effects of asynchronous execution; control simulation should reuse framework-native decision and update rules while preserving closed-loop admission/completion semantics. These motivate \sysname's dual-plane design.

\section{\sysname Design}

\ycrep{This section presents the design of \sysname, which addresses the challenges identified in \S\ref{sec:challenges} and decouples execution modeling from state-dependent serving control. The key insight is that execution can be modeled while preserving its timing effects, whereas state-dependent control behavior can be preserved by emulating only the relevant framework modules instead of replaying the full production runtime. This asymmetry motivates \sysname's dual-plane architecture.}{This section presents the dual-plane architecture of \sysname, which realizes the execution- and control-side design implications identified in \S\ref{sec:challenges}.}
\yongchao{[redundant transition] This opening repeats the same computation--control asymmetry that Section~3 should already derive and summarize. Keep the argument in one place: let Section~3 end with the evidence-derived design requirements, and let Section~4 begin immediately with the concrete architecture that satisfies them. A one-sentence backward reference is sufficient; repeating the insight weakens momentum and delays the design.}

\subsection{Framework Overview} \label{sec:4_1}

\begin{figure*}[t]
  \centering
  \includegraphics[width=\textwidth]{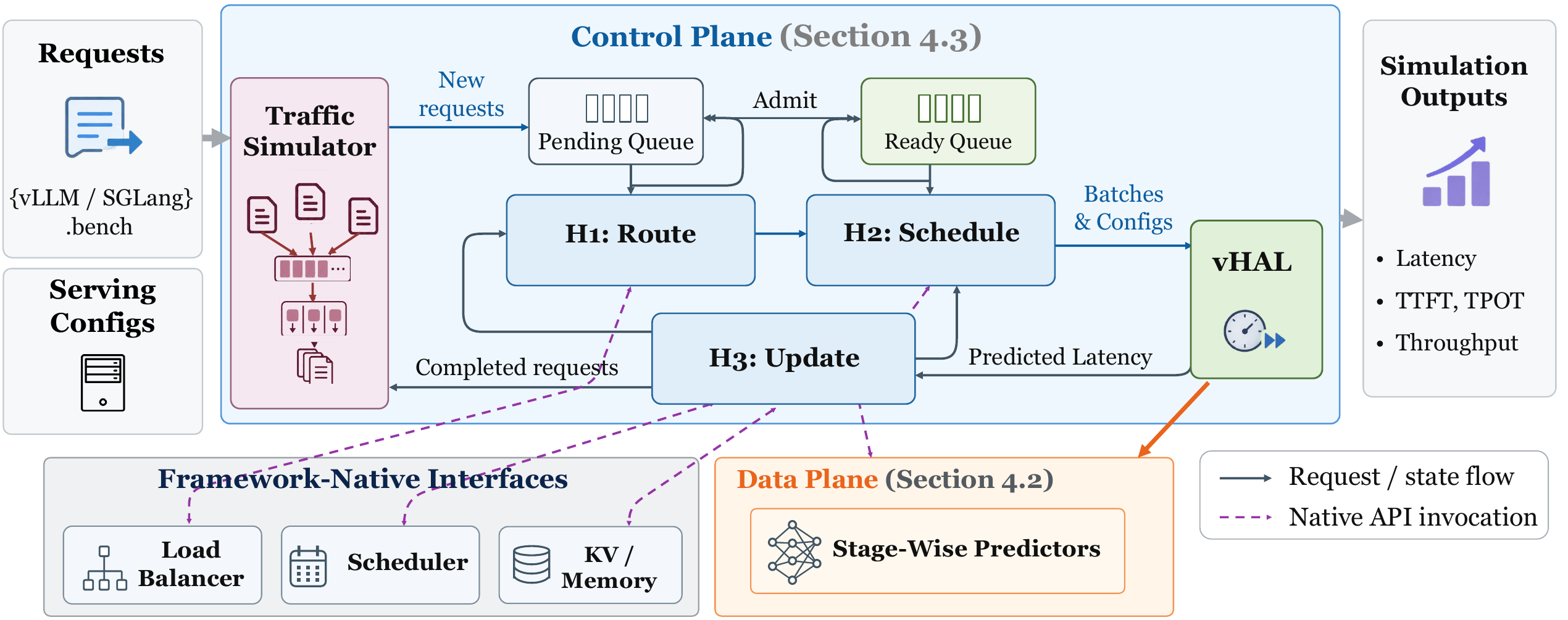}
  \caption{Overview of \sysnames framework. \ycadd{The control plane invokes framework-native control modules over logical state, while the data plane predicts stage latencies; simulated completion drives subsequent control updates through a vHAL interface.}}
  \yongchao{[architecture figure] Figure~1 occupies substantial space but remains a box inventory rather than an explanation of the design: it does not expose the logical-state/event contracts, the vHAL interception boundary, which work is executed versus predicted, or how virtual time and completion events close the loop. Several labels (`Serving SDK,' `Worker Pool,' and `Executor') also do not match the terminology used in the text, and Figure~2 later communicates much of the same high-level flow more clearly. Either remove Figure~1 and enlarge the more informative workflow figure, or redesign Figure~1 around the paper's actual novelty: show the three framework-specific hooks/adapters, the framework-independent state/event interface, predicted stage costs, synthetic completion, and virtual-clock advancement. The caption should state the resulting invariant, not merely say `overview.'}
  \label{fig:framework}
\end{figure*}

\ycrep{
To resolve the speed--accuracy trade-off, \sysname separates serving control from execution timing, as illustrated in Figure~\ref{fig:framework}. The two planes interact iteratively:
}{
Figure~\ref{fig:framework} shows \sysname's dual-plane architecture. The two planes interact iteratively:
}

\begin{itemize}[leftmargin=*, label=$\ast$, itemsep=0pt, topsep=0pt, parsep=0pt]
\item \textit{Control Plane} \ycrep{\sysname maintains logical request and resource state, invokes framework-native control logic at module granularity. A vHAL replaces physical execution with predicted latency, while traffic simulation coordinates request admission (\S4.3).}{maintains logical request and resource state and invokes framework-native control modules. Its vHAL replaces physical execution with predicted completion events on the virtual timeline (\S4.3).}
\item \textit{Data Plane} \ycrep{Given an execution batch and deployment configuration, the data plane predicts its latency using a stage-wise performance model (\S\ref{sec:4_2}) and returns the result to the vHAL.}{predicts pre-forward, forward, and post-forward timing for each execution batch (\S\ref{sec:4_2}).}
\end{itemize}

\ycadd{Native control modules and vHAL execute in wall-clock time to obtain control decisions and predicted delays; their runtime is not added to simulated serving time.
The virtual clock advances to timestamps in the global event queue,
including workload arrivals and predicted completion events.}

\yongchao{[core-boundary clarity] The dual-plane boundary is still operationally ambiguous: the data plane predicts pre/post CPU intervals that include preparation and completion, while the control plane executes native scheduling/update modules. Add a small table stating which production-time costs are predicted, which modules run only to obtain decisions, and why their wall-clock cost is not double-counted.}

\subsection{Stage-Wise Performance Modeling} \label{sec:4_2}

To model iteration latency in asynchronous and potentially
disaggregated serving systems (\S\ref{sec:3_1}), \sysname uses a stage-wise performance model. It represents \ycrep{each iteration with three macroscopic latency targets measured at stable framework boundaries}{\emph{pre-forward}, \emph{forward}, and \emph{post-forward} timing targets from framework-visible signals around model execution}, capturing the resolved effects of CPU--GPU overlap without reconstructing the underlying fine-grained execution.

\subsubsection{Stage Decomposition}
\label{sec:stage_decomposition}
\begin{figure}[t]
  \centering
  \includegraphics[width=\columnwidth]{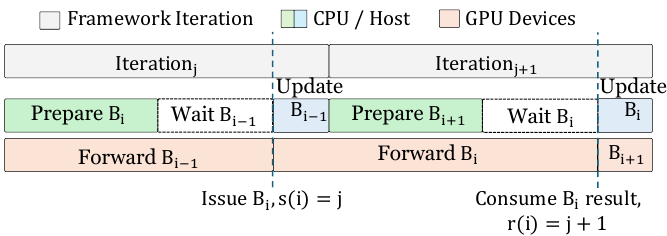}
  \caption{
  Alignment between serving iterations and asynchronous batch execution in the profiled runtime. An iteration may issue a new batch while consuming completion-side effects from an earlier batch.
  }
  \label{fig:stage_alignment}
\end{figure}

Modern serving engines execute an iteration-based loop while model execution proceeds asynchronously. An execution batch formed in serving iteration $j$ may remain active while host-side work for subsequent iterations proceeds, and its completion may be consumed by the framework in a later iteration. Figure~~\ref{fig:stage_alignment} illustrates this mismatch between serving-iteration progress and asynchronous batch execution.

In the profiled runtime, let $s(i)$ denote the serving iteration in which batch $B_i$ is issued and $r(i)$ the iteration in which its completion is consumed by the native update path.
Asynchronous execution permits $r(i)\geq s(i)$, so the batch issued during the iteration need not be the batch whose completion is processed during the same iteration.
During trace construction, \sysname associates each batch's descriptors and timing signals with its issuing iteration, $s(i)$, even when the native update path consumes its completion later.

\sysname aligns batch-associated timing measurements with the
serving-iteration timeline. Let $b_j$ and $b_{j+1}$ mark consecutive
framework-visible scheduling entries of the same scheduler.
The measured iteration latency is
\begin{equation}
L_j=b_{j+1}-b_j.
\label{eq:stage_interval}
\end{equation}

Consider the batch $B_i$ issued in iteration $j=s(i)$.
Let $p_j$ denote the end of host-side batch preparation, and $c_j$ the end of profiled sampling and completion-side processing, excluding detokenization and response dispatch; $c_j$ marks a batch-side measurement boundary distinct from scheduler-side retirement. Here, $F_j$ uses the framework's device-side forward timing and remains associated with the issuing batch. \sysname constructs the three timing targets as

\begin{equation}
\begin{aligned}
T_{\mathrm{fwd},j} &= F_j,\\
T_{\mathrm{post},j} &= (c_j-p_j)-F_j,\\
T_{\mathrm{pre},j} &= L_j-T_{\mathrm{fwd},j}
                          -T_{\mathrm{post},j}.
\end{aligned}
\label{eq:stage_targets}
\end{equation}

\begin{equation}
T_{\mathrm{pre},j}
=
(p_j-b_j)+(b_{j+1}-c_j).
\label{eq:stage_alignment}
\end{equation}
The second term is a signed boundary correction: it subtracts
the portion extending beyond the next scheduling entry, or adds
the gap preceding that entry. Thus, the pre-forward target combines
preparation with the correction needed to align batch-associated
measurements to the scheduler interval.

This construction gives
$T_{\mathrm{pre},j}+T_{\mathrm{fwd},j}+T_{\mathrm{post},j}=L_j$.
Across consecutive intervals,
$\sum_{j=m}^{n}L_j=b_{n+1}-b_m$,
so the target sums account for elapsed time on the scheduler
timeline without accumulating overlapping batch lifetimes.
The stage targets therefore constitute an additive attribution
of iteration latency, rather than three disjoint segments of
a batch's physical execution.

For each stage $s$, \sysname fits a regressor $f_s$ to its measured
target using batch, model, deployment, and platform features
$x_{s,j}$ available at prediction time.
\begin{equation}
\widehat T_{s,j}=f_s(x_{s,j}),
\qquad s\in\{\mathrm{pre},\mathrm{fwd},\mathrm{post}\}.
\label{eq:stage_prediction}
\end{equation}
The predicted serving-iteration latency is
\begin{equation}
\widehat L_j
=
\widehat T_{\mathrm{pre},j}
+
\widehat T_{\mathrm{fwd},j}
+
\widehat T_{\mathrm{post},j}.
\label{eq:stage_sum}
\end{equation}

The vHAL uses $\widehat L_j$ as a logical serving-step duration under the scheduler-boundary approximation. Signed stage targets are retained during training without per-stage clipping; all stage predictions in our evaluation are positive, and hence $\widehat L_j>0$.
\S\ref{sec:stage_ablation} evaluates the stage-wise formulation against direct monolithic prediction.

\subsubsection{Modeling Method} \label{sec:modeling_method}

\ycrep{
Given the stage decomposition above, \sysname proposes a lightweight learning-based method that achieves high fidelity through sparse profiling.
}{
Given the stage targets above, \sysname learns their latencies from real-hardware serving traces. The predictors consume execution-batch, model, deployment, and platform descriptors available at simulation time and must remain inexpensive enough to query throughout a serving simulation.
}

\yongchao{[missing design rationale] This subsection jumps directly from the stage definitions to `a lightweight learning-based method' without explaining why this path follows from the requirements. State the decision criteria first---limited labels, nonlinear regime changes, cross-platform transfer, interpretability, and inference cost---then compare the plausible alternatives and explain which evidence eliminates each one. Also test whether a stage latency is conditionally identifiable from the listed batch/model/platform features: replay matched batches while varying prior-device completion, CPU contention, transfer readiness, and concurrent-worker state, and report $\mathrm{Var}(T_s\mid B)$. If that variance is material, expose the missing state as features or model waiting through explicit resource clocks rather than asking the regressor to absorb hidden queueing state.}

\para{Trace collection.}\ycrep{
To collect ground-truth stage latencies, \sysname instruments the serving framework with lightweight probes at the stage boundaries defined in \S\ref{sec:stage_decomposition} and records $T_{\mathrm{pre}}$, $T_{\mathrm{fwd}}$, and $T_{\mathrm{post}}$ for each serving iteration. During a profiling run, these probes record the wall-clock duration of every iteration, yielding a dataset of \emph{per-iteration, stage-wise} latency measurements under the real framework and hardware.}{
To collect batch-side ground-truth stage latencies, \sysname instruments framework-visible boundaries around model execution and preserves batch identity across asynchronous execution. Probe-on/off runs on vLLM and SGLang show only limited differences in workload E2E duration (Appendix Fig.~\ref{fig:probe_overhead}). The concrete probe locations are backend-specific. We profile 396 configurations (\S\ref{sec:exp_setup}), of which 273 complete successfully and yield usable per-iteration traces for predictor training. The remaining 123 configurations were excluded as infeasible under the evaluated deployment or memory constraints.
}

\ycdel{A small number of representative benchmark runs cover the input space by varying batch size, input and output sequence length (ISL/OSL), and parallelism configuration.}

\yongchao{[missing implementation detail] `A small number' and `representative' are not auditable. Please report the number of runs, configurations, iterations, sampled batch shapes, models, platforms, and frameworks, and state whether the three regressors are global or trained per framework/platform. This information is central to the sparse-profiling claim.}

\para{Regression.}
\ycrep{
Compared to interpolation approaches that require dense grid sampling~\cite{lin2024apex,peng2025reallm}, learning-based regression generalizes performance patterns from sparse data, effectively mitigating the prohibitive costs of exhaustive profiling. \sysname learns the mapping from workload, model, and platform features to the measured stage latency, and employs XGBoost~\cite{chen2016xgboost}, a gradient-boosted decision-tree model as the core regressor to model execution latency. We choose XGBoost for its data efficiency: ground-truth stage latencies require costly real-hardware profiling, while deep neural networks typically demand substantially more training data to realize their advantages. In contrast, tree-based models perform robustly on small tabular datasets~\cite{NEURIPS2021_9d86d83f}, making them well suited to \sysnames limited profiling budget. }{The stage-wise formulation is independent of a particular regression algorithm. Our implementation uses XGBoost~\cite{chen2016xgboost}, which supports nonlinear regression over heterogeneous tabular features and incurs low prediction overhead. We train separate predictor families by framework and stage target.
For a predictor, we pool the profiling iterations and randomly
split them 80\%/20\% into fitting and validation sets; iterations from the same serving run or configuration may appear in both partitions.
}

\yongchao{[why-not-X] The generic tabular-learning citation does not establish that XGBoost is the right choice for these stage-latency traces. Compare it with at least a small MLP and a simple interpolation or linear baseline using the same grouped train/test split, and show error versus the number of serving-trace samples; this would directly support the claimed data efficiency without relying on hypothetical numbers.}

\ycrep{
\para{Feature selection.}
Feature selection is guided by system-level performance factors. We construct a unified feature vector spanning workload, model architecture, and platform characteristics: (1)~\emph{workload features} (batch size, prefill/decode token counts, sequence length statistics), (2)~\emph{model architecture features} (layer count, hidden size, Grouped-Query Attention (GQA) ratio, MoE expert count, MoE routing width, derived FLOP and memory-bandwidth proxies), and (3)~\emph{platform performance features} (measured GEMM throughput, attention kernel efficiency, memory bandwidth, and NCCL latency/bandwidth from microbenchmarks). These features form a compact representation for stage-latency regression across heterogeneous serving configurations.}{
\para{Feature construction and encoding.}
The regressors consume batch, model/deployment, and platform descriptors available at simulation time. Model and platform descriptors remain predictor inputs, allowing each family to span the profiled models and platforms. Derived features encode prefill/decode-specific operation scale, KV-memory footprint, and communication demand. Appendix~\ref{app:features} provides the feature-family schema, provenance, encoding, stage applicability, and platform-descriptor profiling procedure.}

\subsection{Module-Level Control-Plane Emulation} \label{sec:4_3}

\ycrep{
To reduce adaptation overhead as serving policies evolve (\S\ref{sec:3_2}), \sysname preserves serving control at module granularity by executing the framework's native control logic over simulated request and resource state. 
}{
To avoid duplicating state-dependent serving policies in the simulator, \sysname executes framework-native control logic at module granularity over simulated request and resource state.}
A vHAL substitutes the physical execution environment, allowing these modules to run without replaying the surrounding production runtime. 

\begin{figure}[t]
  \centering
  \includegraphics[width=\columnwidth]{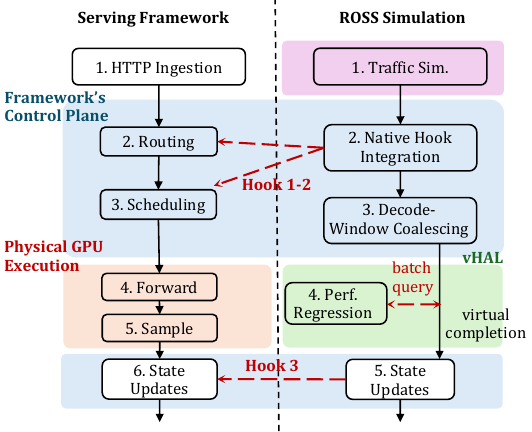}
  \caption{Control-plane data flow in a serving framework (left) and \sysname (right). \sysname \ycrep{preserves native control logic at module granularity while replacing physical execution with predicted stage latencies.}{reuses routing, scheduling, and state-update modules through three hooks, while vHAL replaces physical execution with predicted virtual completion.}}
  \label{fig:workflow}
\end{figure}

Figure~\ref{fig:workflow} contrasts the \ycrep{control-plane data flow}{execution boundary} of a serving framework with that of \sysname. 
\ycrep{
In the serving framework, admitted requests are routed and scheduled into execution batches; after physical execution, the returned results drive subsequent state updates. In \sysname, traffic simulation supplies ready requests; native control modules use the current logical state to form execution batches; decode-window coalescing optionally groups stable decode iterations; the performance model predicts their stage latencies; and the vHAL advances virtual time before the update path advances logical state. Within each reused module, \sysname executes the same control paths as the serving framework, while omitting the surrounding runtime services and physical workers.
}{
\sysname retains framework-native routing, scheduling, and state-update logic, while the vHAL replaces physical model execution with predicted virtual completion; the traffic controller supplies request admissions. Algorithm~\ref{alg:sim_loop} specifies how these components are composed over virtual time. A worker is runnable when it has schedulable work and no outstanding vHAL submission.
}

\para{vHAL \ycadd{and virtual completion}.}
\ycrep{The vHAL provides the execution interface required by the native control modules without invoking physical model execution. During initialization, it replaces device-dependent setup such as weight loading and memory allocation with lightweight logical state.
At runtime, it intercepts each execution call, extracts the batch descriptor $B$ and queries the performance model $\mathcal{M}.\textsc{Predict}(B)$ for $(T_{\mathrm{pre}}, T_{\mathrm{fwd}}, T_{\mathrm{post}})$. The virtual clock is then advanced by the predicted latency, after which the native update path proceeds as if the corresponding forward pass had completed.
Thus, each ordinary iteration requires only a fixed number of lightweight model queries and no physical GPU execution.
}{\sysname uses a scheduler-boundary approximation. For a batch $\mathcal B$ scheduled by worker $w$ at virtual time $t$, the vHAL predicts a logical-step duration $\widehat L$ and queues a \textsc{VirtualComplete} event at $t+\widehat L$. When the event is processed, the native update path applies completion-side request and KV-state updates for $\mathcal B$ before $w$ makes its next scheduling decision.
Thus, each worker is simulated as a sequence of logical schedule--update steps rather than an explicit in-flight batch pipeline.
}

\yongchao{[vHAL design reproducibility] This reads as an implementation sketch, not a reproducible abstraction. Define the vHAL contract formally: the calls/events it intercepts; the framework-independent batch, resource, and completion state it exposes; the synthetic outputs returned to native code; and the ordering semantics for asynchronous GPU work, collectives, KV transfers, callbacks, and virtual-clock advancement. Then separate the common vHAL state/event machine from framework-specific adapters and provide a table for vLLM, SGLang, and Dynamo mapping each native API/object to that contract, including initialization, device properties, memory/KV capacity, model-executor calls, and completion/update callbacks. A worked trace should show one request from admission through interception, prediction, synthetic completion, state mutation, and the next schedulable event. Without this, readers know that an interception layer must exist but cannot determine how heterogeneous frameworks share one simulation core or reproduce it. Also replace the implicit $O(1)$ claim with the precise complexity in tree count/depth, number of stage queries, workers, and decode-window length.}

\para{Native control invocation.}
The reused control logic operates on logical request and resource
state rather than physical tensor values. \sysname maintains the state expected by the framework and invokes its native control logic through three narrow hook points, corresponding to the \texttt{[Hook]} annotations in Algorithm~\ref{alg:sim_loop}:

\begin{itemize}[leftmargin=*, label=$\ast$, itemsep=2pt, topsep=2pt, parsep=0pt]
\item \textbf{Hook 1 ($\Pi_{\mathrm{LB}}.\textsc{Route}$):} \ycrep{The native routing logic assigns an admitted request to a worker pool using the current per-worker load state.}{Given an admitted request and current per-worker load state, the native routing logic selects the target worker pool.}
\item \textbf{Hook 2 ($\Pi_{\mathrm{SCHED}}.\textsc{Schedule}$):} \ycrep{The native scheduler consumes the current request and KV-cache state and produces an execution batch $B$.}{Given the current request and KV/block state, the native scheduler determines admissions, preemptions, memory allocations, and the next execution batch $B$.}
\item \textbf{Hook 3 ($\Pi_{\mathrm{SCHED}}.\textsc{Update}$):}\ycrep{After virtual time advances, the native update routine advances request and KV-cache state following the completed execution step.}{ Given a logical completion event, the native update path advances request progress and applies the corresponding KV/block state changes.}
\end{itemize}

\ycadd{\sysname maintains framework-compatible request, KV/block, and worker state while supplying request arrivals, virtual time, and vHAL completions; native modules determine the corresponding control transitions. Appendix~\ref{app:native_control} details the reused modules, implementation footprint, and framework-visible control state.}

\ycadd{
Under the scheduler-boundary approximation, native-control reuse gives a checkable per-invocation consistency condition: given the same native policy, equivalent hook-visible pre-state and identical hook inputs, native and simulated invocations should produce the same exposed decisions and state updates.
\S\ref{sec:control_fidelity} evaluates this condition under a preemption-policy change. Policy changes confined to the reused modules are inherited through the same hooks; changes to hook-visible state or interface semantics may require corresponding adapter updates.
}

\yongchao{[native-module contract] Naming three hooks does not explain how native modules remain semantically valid outside their original runtime. Specify the preconditions and postconditions of each hook and ownership of request/KV/block/worker state. Then define and test a state-equivalence invariant against real execution---for the same arrivals and modeled completion times, the native and simulated paths should produce the same ordered control decisions and resource-state transitions. This is the missing design argument between `reuse native modules' and `therefore preserve fidelity.'}

\yongchao{[adaptation claim] Three hook points do not by themselves establish low adaptation overhead. Report the \emph{\sysname-side} code changes and engineer effort required for the modified preemption policy and for at least one framework-version upgrade; this directly tests whether native reuse actually reduces simulator maintenance.}

\begin{algorithm}[t]
\caption{\sysname event loop across serving pools}
\label{alg:sim_loop}
\begin{algorithmic}[1]
\REQUIRE Traffic controller $\mathcal{A}$;
native load balancer $\Pi_{\mathrm{LB}}$;
schedulers $\{\Pi^{w}_{\mathrm{SCHED}}\}_{w\in\mathcal{W}}$;
vHAL $\mathcal{V}$

\STATE Initialize event queue $\mathcal{Q}$ and logical states
$\{\Sigma_w\}_{w\in\mathcal{W}}$

\WHILE{unfinished requests remain}
    \STATE $(t,\mathcal{E}_t) \gets \mathcal{Q}.\textsc{PopNext}()$
    \STATE $\mathcal{W}_t \gets \emptyset$

    \FOR{each event $e \in \mathcal{E}_t$}
        \IF{$e=\textsc{VirtualComplete} (w,\mathcal B,k,\boldsymbol{\tau})$}
            \STATE \textbf{[Hook 3]}
            $\Delta \gets \Pi_{\mathrm{SCHED}}^w.\textsc{Update}
                (\Sigma_w,\mathcal B,k,\boldsymbol{\tau})$
            \STATE $\mathcal A.\textsc{OnFinished}
                (\Delta.\mathrm{finished},t)$
            \STATE Apply prefill-to-decode handoff, if any, and add
                newly runnable destination workers to $\mathcal W_t$
            \STATE $\mathcal W_t \gets \mathcal W_t \cup \{w\}$
        \ELSE
            \STATE $\mathcal A.\textsc{Apply}(e,t)$
        \ENDIF
    \ENDFOR
    
    \FOR{each $r \in \mathcal{A}.\textsc{AdmitReady}(t)$}
        \STATE \textbf{[Hook 1]}
        $w \gets \Pi_{\mathrm{LB}}.\textsc{Route}(r)$
        \STATE Enqueue $r$ at worker $w$;
        $\mathcal{W}_t \gets \mathcal{W}_t \cup \{w\}$
    \ENDFOR

    \FOR{each runnable $w \in \mathcal{W}_t$}
        \STATE \textbf{[Hook 2]}
        $\mathcal{B} \gets
        \Pi^{w}_{\mathrm{SCHED}}.\textsc{Schedule}(\Sigma_w)$

        \IF{$\mathcal{B} \neq \emptyset$}
            \STATE $k \gets
            \textsc{DecodeWindow}(w,\mathcal{B},t,\mathcal{Q})$
            \STATE $\mathcal{Q} \gets
            \mathcal{Q} \cup
            \mathcal{V}.\textsc{Submit}(w,\mathcal{B},k,t)$
        \ENDIF
    \ENDFOR
\ENDWHILE
\end{algorithmic}
\end{algorithm}

\para{Traffic simulation.}
\ycrep{
Production API gateways typically cap in-flight requests via a concurrency semaphore, creating a feedback loop between client pacing and server scheduling. \sysname models this with a \emph{Semaphore Controller} operating entirely on the virtual timeline. When the concurrency limit is reached, newly ready requests remain pending until a simulated completion frees capacity, at which point they become eligible for admission. This avoids physical blocking while preserving the back-pressure and queueing behavior. Unlike full-runtime time-warp execution, this
design does not require blocked clients or workers to remain active on the wall-clock timeline.
}{
The traffic controller $\mathcal A$ maintains client-side request readiness, admission, and concurrency state on the same virtual timeline as server-side control. This separation is deliberate: $\mathcal A$ determines \emph{when} a request may enter the serving system, while framework-native routing and scheduling determine \emph{where} it is placed and \emph{what} executes next. \S\ref{sec:traffic_ablation} evaluates these closed-loop admission--completion dynamics against physical execution.
}

\ycadd{
Under a concurrency limit, a request completion releases its in-flight slot and may admit a pending request at the same virtual time. Request arrival times and output lengths are supplied by the workload; a request terminates after generating its specified output length.
}
\yongchao{[traffic-simulation design] This paragraph states the desired effect but not the simulation design. Define the workload and event model: supported trace and synthetic inputs; open-loop arrivals versus closed-loop clients with think time; global versus per-client concurrency caps; queued, admitted, executing, streaming, and completed request states; and exactly which completion releases a semaphore slot. State whether output length and stop events are oracle trace fields, sampled before execution, or generated causally from simulated outputs. Specify the event set (arrival, admission, forward/iteration completion, KV-transfer completion, request completion, and deferred callback), event-queue data structure, timestamp tie-breaking/priority rules, and interactions with multiple workers or PD pools. Provide pseudocode showing how arrivals and completions create future events and how the controller communicates with native routing/scheduling. Then validate admission timestamps, outstanding-request trajectories, queue length, throughput, and tail latency against an instrumented real gateway over a concurrency-cap sweep. The currently disabled appendix lists workload generators, but it still does not supply these semantics.}

\para{Simulation event loop.}
\ycdel{Algorithm~\ref{alg:sim_loop} integrates virtual-time traffic admission, native control invocation, and predictive execution. At each iteration, the traffic controller admits requests ready at the current virtual time, after which the native control logic forms an execution batch. The performance model predicts its stage latencies, the virtual clock advances accordingly, and the native update path advances the corresponding request and resource state.
}\ycadd{Algorithm~\ref{alg:sim_loop} composes traffic, native control, and virtual execution through a global event queue. \textsc{VirtualComplete} denotes a logical serving-step completion. At each virtual timestamp, state-changing events are applied before new admissions and scheduling decisions; vHAL submissions insert future completion events into the same queue. Data-parallel workers are modeled as independent replicas. For PD serving, \sysname models KV transfer as an instantaneous logical handoff at prefill completion, after which the request becomes eligible for decode scheduling subject to KV/resource availability.}
\yongchao{[event semantics] The empty-batch branch in Algorithm~\ref{alg:sim_loop} advances only to the next arrival, but PD transfer completion or another deferred state transition may unblock scheduling first. Please define whether such events are folded into \textsc{NextArrivalTime}, or replace this with a general \textsc{NextEventTime}.}

\ycdel{
To reduce redundant control invocations during stable decode phases, \sysname applies a decode-window coalescing mechanism.
}
\yongchao{[why-not-X] The mechanism is described, but its contribution is not measured. Add an ablation with coalescing disabled and report scheduler invocations, model-query time, total simulation runtime, and any fidelity difference across short- and long-output workloads.}
\ycdel{
After an ordinary decode iteration, \textsc{DecodeWindow} identifies the longest interval over which the active batch and scheduling outcome remain unchanged, ending before the next arrival, request completion, or other scheduling event.
}
\ycrep{For a window of $k>0$ iterations, the performance model batches their features into a single XGBoost query and predicts the latency sequence $L_{1:k}$, and \textsc{AdvanceDecode} applies the corresponding request and KV-state updates in order without reinvoking the scheduler.}{
After Hook~2 forms a decode batch $\mathcal B$, \textsc{DecodeWindow} selects $k$ logical steps until the next admission, request completion or KV/resource boundary that may change scheduling. Candidate windows are bounded by running/KV state and truncated using predicted step endpoints.
\sysname evaluates their per-step features with one batched query per stage. Starting at virtual time $t$, the step endpoints are
\[
\tau_\ell =
t+\sum_{q=0}^{\ell-1}\widehat L_{j+q},
\qquad \ell=1,\ldots,k.
\]
Completion-side updates and token timestamps follow these endpoints in logical-step order; native scheduling resumes at $\tau_k$. \S\ref{sec:coalescing_ablation} evaluates the resulting runtime benefit.
}


\section{Evaluation} \label{sec:evaluation}

\ycadd{We organize the evaluation around five questions:
(1) How accurately does \sysname reproduce end-to-end serving performance?
(2) How do its core design components affect fidelity and efficiency?
(3) How robust is \sysname across workloads, models, and long-context serving regimes?
(4) How fast and scalable is \sysnames simulation?
(5) How does \sysname perform in representative operational scenarios? The final question is examined through two case studies.}

\yongchao{[evaluation spine] Add an explicit roadmap of research questions before the setup: overall predictive accuracy, benefit of stage-wise modeling, native-control fidelity, closed-loop traffic correctness, generalization, simulation efficiency, and operational utility. Map each question to one subsection and one design claim so readers can audit claim coverage.}

\subsection{Experiment Setup} \label{sec:exp_setup}

\para{Testbed and models.}
We evaluate on NVIDIA Blackwell B200 and Hopper H200 GPUs using open-weight LLMs spanning scales and architectures: \textit{Llama-3.1}~\cite{llama2024llama3} (8B, 70B), \textit{Qwen-2.5/3}~\cite{yang2024qwen25,qwen2025qwen3} (72B, 30B-A3B, 32B, 235B-A22B), and \textit{DeepSeek-V3}~\cite{liu2024deepseek,deepseek_v3_0324_hf} (671B MoE, official FP8 checkpoint).
\ycadd{Generalization experiments add \textit{QwQ-32B}~\cite{qwq32b} (dense) and \textit{GPT-OSS-120B}~\cite{openai2025gptoss120bgptoss20bmodel} (MoE) as calibration-held-out models, while \S\ref{sec:stage_ablation} uses held-out Qwen3-14B~\cite{qwen2025qwen3}.}

We use vLLM (v0.11.2) and SGLang (v0.5.6). Disaggregated vLLM deployments use NVIDIA Dynamo (v0.7.0)~\cite{nvidia2025dynamo} for prefill–decode routing and KV-cache transfer. \ycrep{Unless noted otherwise, SGLang fidelity results use the patched version to isolate the serving paths modeled by \sysname; \S\ref{sec:patching} separately evaluates the stock post-generation bottleneck.}{SGLang measurements use vanilla v0.5.6, with the TokenizerManager patch applied to \S\ref{sec:generalization} runs with maximum batch size $\ge128$; \S\ref{sec:patching} compares vanilla and patched versions.
}

\yongchao{[target validity] Most SGLang fidelity numbers use a patched system that changes an unmodeled production path, yet the headline presents them as accuracy on SGLang. Make stock-SGLang results primary, retain patched results as an isolation ablation, and specify the exact patch and affected measurements.}

\para{Datasets and workloads.}
\ycrep{We use ShareGPT~\cite{sharegpt2026}
to emulate production workloads, sweeping input/output sequence lengths (\textit{ISL/OSLs}) from 500 to 2500 tokens to stress system dynamics. Dataset-distribution ablations additionally use RepoQA~\cite{repoqa} (long-context code understanding) and AIME 2026~\cite{aime26} (mathematical reasoning) at ISL/OSL$=$2500/2500.}{We use ShareGPT~\cite{sharegpt2026} as the primary workload source. RepoQA~\cite{repoqa} and AIME 2026~\cite{aime26} are used for dataset-robustness experiments.}
\ycadd{Table~\ref{tab:eval_overview} summarizes the evaluation setup; complete settings are reported in Appendix Table~\ref{tab:eval_setup_full}.}

\ycadd{
\para{Metrics and evaluation data.}
We report workload E2E duration as benchmark duration and TPOT as the mean of request-level TPOT values. We compute absolute percentage error (PE) as
$\mathrm{PE}(m)=|\hat{m}-m|/m\times100\%$,
where $m$ and $\hat{m}$ denote the measured and simulated values, respectively. Fidelity results use separately collected serving runs, distinct from the profiling traces used to train the stage-wise predictors (\S\ref{sec:modeling_method}), while their configurations may overlap the profiling set; held-out conditions are specified in the corresponding subsections.
}

\ycdel{
\para{Profiling dataset.}
The stage-wise regressors (\S\ref{sec:4_2}) are trained on per-iteration latencies collected from real execution traces under varied parallelism configurations, ISL/OSL combinations, and batch sizes. We pool the iteration samples from all profiled runs and randomly split them 80\%/20\% into training and validation sets. This split is used only for fitting and within-profile validation.}

\yongchao{[leakage] Randomly splitting pooled iterations lets highly correlated samples from the same run and configuration appear in both fitting and validation. Use a grouped split by complete run/configuration for within-profile validation. Separately disclose exactly which headline table cells overlap the profiling conditions, so interpolation results are not mistaken for held-out generalization.}
\ycdel{Tables~\ref{tab:fidelity_baseline} and~\ref{tab:fidelity_ross} report overall end-to-end fidelity and may include conditions represented in profiling. Generalization to held-out models, datasets, and workload conditions is evaluated separately in \S\ref{sec:generalization}.}

\para{Simulation baselines.}
\yongchao{[baseline fairness] \sysname may see target conditions during profiling, while Vidur is moved to unsupported hardware and sometimes uses stale framework profiles. Add a native-supported comparison or give every simulator the same hardware/framework version and calibration budget; otherwise these errors cannot isolate architectural differences.}
\ycrep{We compare \sysname with the open-source serving simulators Vidur~\cite{agrawal2024vidur} and AIConfigurator~\cite{aiconfigurator2026} with minimal compatibility patches. Vidur natively targets H100 and supports neither B200 nor PD-disaggregated deployment; we apply a minimal patch to run it on H200, feasible because H100 and H200 share similar microarchitecture. AIConfigurator supports more models, but for consistency with Vidur, Table~\ref{tab:fidelity_baseline} evaluates both on the same subset: Llama-3.1-70B, Qwen3-32B, Qwen2.5-72B, and DeepSeek-V3.}{We compare \sysname with the open-source serving simulators Vidur~\cite{agrawal2024vidur} and AIConfigurator~\cite{aiconfigurator2026} on H200, applying minimal compatibility patches where required. We evaluate each simulator on its available configurations (Table~\ref{tab:fidelity_baseline}).}

\ycdel{
\para{Default configuration for generalization tests.}
Unless otherwise noted, experiments (\S\ref{sec:ablation}) use B200, Llama-3.1-70B, parallelism data-parallel (DP) : pipeline-parallel (PP) : tensor-parallel (TP) $=$ 1:1:8, ISL/OSL$=$2500/2500, and a saturated request rate with 512 prompts. Unseen-model experiments additionally evaluate \textit{QwQ-32B}~\cite{qwq32b} (dense) and \textit{GPT-OSS-120B}~\cite{openai2025gptoss120bgptoss20bmodel} (MoE) on both vLLM and SGLang.
}
\yongchao{[consistency] The abstract, introduction, and conclusion say `seven models,' but this subsection adds two models held out from calibration. Use one precise formulation throughout, e.g., `seven main-evaluation models plus two calibration-held-out models'; this also makes the generalization evidence visible in the headline.}

\begin{table*}[t]
  \centering
  \caption{
  Evaluation overview. Complete per-experiment settings are reported in Appendix Table~\ref{tab:eval_setup_full}. Unless otherwise noted, we use each backend's native memory-fraction setting at 0.9 and a chunked-prefill size of 8192 tokens (Appendix~\ref{app:common_settings}).
  }
  \label{tab:eval_overview}

  \small
  \setlength{\tabcolsep}{4pt}
  \renewcommand{\arraystretch}{1.08}

  \begin{tabularx}{\textwidth}{@{}
    >{\raggedright\arraybackslash}p{.2\textwidth}
    >{\raggedright\arraybackslash}p{.28\textwidth}
    >{\raggedright\arraybackslash}X
    @{}}
    \toprule
    Experiment & Coverage & Key Setup \\
    \midrule

    \S\ref{sec:e2e} Overall fidelity
    & Seven main-evaluation models; vLLM/SGLang; H200/B200;
      co-located and PD-disaggregated serving
    & Prompt/generation-length mixes and parallelism configurations
    \\

    \S\ref{sec:stage_ablation} Stage decomposition
    & Dense-model traces on H200; vLLM
    & Stage-wise vs.\ monolithic latency modeling
    \\

    \S\ref{sec:control_fidelity} Native control
    & Qwen3-32B; vLLM; H200
    & Modified preemption policy; scheduler decisions and KV-cache state
    \\

    \S\ref{sec:traffic_ablation} Closed-loop traffic
    & Qwen3-32B; vLLM; H200
    & Finite-rate vs.\ saturated traffic under concurrency-limited admission
    \\
    
    \S\ref{sec:coalescing_ablation} Decode coalescing
    & Qwen2.5-72B-Instruct; vLLM; H200
    & Coalescing on/off under short- and long-generation workloads
    \\
    
    \S\ref{sec:prefill} Long-context
    & Qwen3-235B-A22B; SGLang; B200
    & Context length and prefix-cache reuse
    \\

    \S\ref{sec:generalization} Generalization
    & Llama-3.1-70B (held-out: QwQ-32B and GPT-OSS-120B);
      vLLM/SGLang; B200
    & Input/output sequence length (ISL/OSL), request-rate, batch-size, dataset, and calibration-held-out-model sweeps
    \\

    \S\ref{sec:scalability} Simulation scaling
    & Llama-3.1-70B; vLLM; B200
    & CPU-core budget and concurrent simulation instances
    \\

    \S\ref{sec:use_case} Configuration search
    & Qwen3-32B; SGLang; B200
    & Co-located/disaggregated search over 228 candidate configurations
    \\

    \S\ref{sec:patching} Performance diagnosis
    & Qwen3-32B; SGLang; B200
    & Maximum-batch-size sweep; vanilla vs.\ patched SGLang
    \\

    \bottomrule
  \end{tabularx}
\end{table*}

\subsection{End-to-End Simulation \ycrep{Performance}{Fidelity}} \label{sec:e2e}
\yongchao{[claim×experiment gap] The paper's central design contribution is the stage-wise decomposition (§4.2), yet no experiment isolates its benefit over a monolithic (single-stage) regressor. A reviewer will ask: `What if you predicted $L_i$ directly instead of $T_{\mathrm{pre}} + T_{\mathrm{fwd}} + T_{\mathrm{post}}$?' Add an ablation: train a monolithic XGBoost regressor on the same data predicting $L_i$ directly, and compare PE. If the stage-wise approach is genuinely better at capturing overlap, this should be demonstrable with one experiment. Without it, the decomposition is an unsubstantiated design choice.}

\ycrep{
We first evaluate overall end-to-end fidelity across hardware platforms, frameworks, models, and deployment modes. For each model--deployment--platform cell, we vary three ISL/OSL distributions and three parallelism configurations for co-located deployments (or two for disaggregated) at a fixed request rate and batch size.
}{
For each model--GPU--backend--deployment cell, we evaluate all applicable configurations summarized in Table~\ref{tab:eval_setup_full} and report the mean configuration-level PE. Headline statistics summarize the distribution of these cell-level means.
}
\yongchao{[setup completeness] Please enumerate the three ISL/OSL distributions, all parallelism configurations, and the fixed request rate and batch-size limit. The cell means and the claimed 52-cell coverage are otherwise not reproducible.}
\ycdel{ Tables~\ref{tab:fidelity_baseline} and~\ref{tab:fidelity_ross} report the \emph{mean} PE within each cell; the headline statistics below summarize the distribution of these cell-level means.
} \yongchao{[metric definition] Define PE with a formula and specify whether it is absolute or signed. Also define E2E latency's aggregation unit (per-request latency or workload makespan) and whether each table entry is mean-of-errors or error-of-means; these choices can materially change the values.}
\yongchao{[aggregation] Averaging configurations within each cell before computing the headline median can hide large failures and gives equal weight to cells containing different numbers of settings. Report the full configuration-level distribution, including p90/worst-case error, and justify the hierarchical weighting.}

\begin{table*}[t]
\centering
\caption{%
  \ycrep{Baseline comparison (H200, mean PE [\%])}{End-to-end fidelity on H200 across configurations supported by each
simulator (mean PE [\%]). Each entry is the arithmetic mean of configuration-level PEs within the corresponding model--framework--deployment cell}. \ycdel{\colorbox{gray!15}{Shaded rows} = \sysname (ours).}
  \texttt{N/A} = \ycrep{deployment mode not supported by the baseline}{unsupported or infeasible deployment mode};
  \texttt{---} = \ycrep{model not supported by the baseline}{unavailable model--backend combination}.%
}

\label{tab:fidelity_baseline}
\setlength{\tabcolsep}{4pt}
\small
\begin{tabular}{l l l rr rr rr rr}
\toprule
& & &
\multicolumn{4}{c}{vLLM} &
\multicolumn{4}{c}{SGLang} \\
\cmidrule(lr){4-7} \cmidrule(lr){8-11}
Model & Simulator & &
\multicolumn{2}{c}{E2E Dur.\ PE} & \multicolumn{2}{c}{TPOT PE} &
\multicolumn{2}{c}{E2E Dur.\ PE} & \multicolumn{2}{c}{TPOT PE} \\
\cmidrule(lr){4-5}\cmidrule(lr){6-7}\cmidrule(lr){8-9}\cmidrule(lr){10-11}
& & & Co-loc & Disagg & Co-loc & Disagg & Co-loc & Disagg & Co-loc & Disagg \\
\midrule

\multirow{3}{*}{Llama-3.1-70B}
  & \cellcolor{gray!15}\sysname (Ours) & \cellcolor{gray!15} & \cellcolor{gray!15}3.72 & \cellcolor{gray!15}8.06 & \cellcolor{gray!15}2.19 & \cellcolor{gray!15}2.92 & \cellcolor{gray!15}4.84 & \cellcolor{gray!15}3.15 & \cellcolor{gray!15}7.92 & \cellcolor{gray!15}7.42 \\
& Vidur        & & 39.9 & N/A  & 16.9 & N/A  & ---  & ---  & ---  & ---  \\
& AIConf.      & & 63.9 & 67.2 & 11.4 & 24.9 & 84.0 & 80.9 & 104.7 & 71.8 \\
\midrule

\multirow{3}{*}{Qwen3-32B}
  & \cellcolor{gray!15}\sysname & \cellcolor{gray!15} & \cellcolor{gray!15}1.37 & \cellcolor{gray!15}5.05 & \cellcolor{gray!15}1.41 & \cellcolor{gray!15}4.54 & \cellcolor{gray!15}5.57 & \cellcolor{gray!15}1.35 & \cellcolor{gray!15}5.90 & \cellcolor{gray!15}2.46 \\
& Vidur   & & 172.2 & N/A  & 288.6 & N/A  & ---  & ---  & ---  & ---  \\
& AIConf. & & 70.3  & 97.0 & 27.6  & 26.5 & 87.4 & 83.7 & 90.5 & 72.0 \\
\midrule

\multirow{3}{*}{Qwen2.5-72B}
  & \cellcolor{gray!15}\sysname & \cellcolor{gray!15} & \cellcolor{gray!15}2.86 & \cellcolor{gray!15}7.62 & \cellcolor{gray!15}1.68 & \cellcolor{gray!15}2.98 & \cellcolor{gray!15}3.45 & \cellcolor{gray!15}3.47 & \cellcolor{gray!15}2.87 & \cellcolor{gray!15}4.14 \\
& Vidur   & & 179.5 & N/A  & 250.3 & N/A  & ---  & ---  & ---  & ---  \\
& AIConf. & & 62.1  & 67.2 & 10.7  & 24.2 & 88.9 & 80.7 & 107.2 & 69.5 \\
\midrule

\multirow{2}{*}{DeepSeek-V3}
  & \cellcolor{gray!15}\sysname & \cellcolor{gray!15} & \cellcolor{gray!15}8.49 & \cellcolor{gray!15}N/A & \cellcolor{gray!15}6.30 & \cellcolor{gray!15}N/A & \cellcolor{gray!15}2.02 & \cellcolor{gray!15}N/A & \cellcolor{gray!15}2.80 & \cellcolor{gray!15}N/A \\
& Vidur & & 38.8 & N/A & 81.3 & N/A & --- & --- & --- & --- \\
\bottomrule
\end{tabular}
\end{table*}

\begin{table*}[t]
\centering
\caption{%
  \ycrep{\sysname fidelity across all platforms and models}{Additional \sysname fidelity results beyond Table~\ref{tab:fidelity_baseline}} (mean PE [\%]).
  \ycdel{N/A = deployment mode not evaluated.}
  \ycrep{Models in this table include those unsupported by any baseline and all B200 results (no baseline supports B200).}{The table includes all B200 results and H200 model configurations without baseline coverage.}
  \ycrep{DeepSeek-V3 disaggregated is marked N/A because it exceeds the 8-GPU KV-cache budget.}{N/A = infeasible under the evaluated 8-GPU KV-cache budget.}%
}

\label{tab:fidelity_ross}
\setlength{\tabcolsep}{4pt}
\small
\begin{tabular}{l l rr rr rr rr}
\toprule
& &
\multicolumn{4}{c}{vLLM} &
\multicolumn{4}{c}{SGLang} \\
\cmidrule(lr){3-6} \cmidrule(lr){7-10}
Platform & Model &
\multicolumn{2}{c}{E2E Dur.\ PE} & \multicolumn{2}{c}{TPOT PE} &
\multicolumn{2}{c}{E2E Dur.\ PE} & \multicolumn{2}{c}{TPOT PE} \\
\cmidrule(lr){3-4}\cmidrule(lr){5-6}\cmidrule(lr){7-8}\cmidrule(lr){9-10}
& & Co-loc & Disagg & Co-loc & Disagg & Co-loc & Disagg & Co-loc & Disagg \\
\midrule

\multirow{3}{*}{H200}
& Llama-3.1-8B     & 3.88  & 5.80  & 4.40  & 6.58  & 10.05 & 5.21  & 9.78  & 7.98  \\
& Qwen3-30B-A3B    & 3.24  & 5.54  & 4.02  & 4.61  & 4.44  & 1.64  & 4.50  & 0.94  \\
& Qwen3-235B-A22B  & 0.95  & 8.59  & 0.79  & 8.86  & 0.65  & 0.71  & 1.10  & 0.54  \\
\cmidrule{1-10}

\multirow{7}{*}{B200}
& Llama-3.1-70B    & 5.34  & 8.47  & 5.76  & 5.07  & 5.53  & 7.19  & 8.22  & 8.39  \\
& Llama-3.1-8B     & 3.23  & 8.07  & 4.32  & 7.47  & 7.70  & 9.14  & 5.64  & 14.18 \\
& Qwen3-32B        & 1.32  & 12.06 & 0.52  & 3.64  & 5.44  & 7.73  & 5.29  & 7.02  \\
& Qwen2.5-72B      & 1.39  & 7.44  & 0.83  & 4.33  & 6.74  & 6.50  & 5.78  & 6.09  \\
& Qwen3-30B-A3B    & 4.62  & 8.18  & 4.21  & 1.60  & 5.62  & 1.67  & 5.79  & 2.96  \\
& Qwen3-235B-A22B  & 4.81  & 9.02  & 4.71  & 4.79  & 1.76  & 0.92  & 2.26  & 0.70  \\
& DeepSeek-V3      & 5.44  & N/A   & 2.93  & N/A   & 5.39  & N/A   & 7.09  & N/A   \\
\bottomrule
\end{tabular}
\end{table*}

\para{End-to-end duration and TPOT.} 
Across 52 evaluated \ycdel{model--platform--framework--deployment }cells, \sysname achieves a \textbf{median E2E PE of \ycrep{4.8}{5.3}\% \ycdel{($\sigma$\,=\,\ycrep{5.1}{2.8}\%)}} and \textbf{median TPOT PE of \ycrep{3.9}{4.5}\%\ycdel{($\sigma$\,=\,\ycrep{4.5}{2.8}\%)}}. Considering both metrics yields \ycrep{104 error entries: their combined p90 is 8.3\%, and only 3 of 104 exceed 10\%}{104 cell-level mean errors, with a combined p90 of 8.3\%; only 3 of 104 exceed 10\%. This cell-level summary weights each model--platform--framework--deployment context equally. Across the 273 configuration-level evaluation records, E2E PE has a median and p90 of 3.9\% and 12.4\%, respectively; the corresponding TPOT values are 3.6\% and 11.0\%.}
\yongchao{[must-fix numerical consistency] I cannot reproduce the headline medians or dispersions from the 52 E2E and 52 TPOT cells displayed in Tables~\ref{tab:fidelity_baseline}--\ref{tab:fidelity_ross}. Using each unique displayed \sysname cell gives E2E median 5.275\% and population $\sigma\approx2.78\%$, and TPOT median 4.45\% and population $\sigma\approx2.76\%$; only the combined p90 (8.3\%) and 3/104 count match. Recompute from the authoritative raw data, state the exact population and statistic behind $\sigma$, and distinguish c\sysname-cell dispersion from repeated-run uncertainty with repetition counts and confidence intervals.}
\ycrep{Baselines drift severely because they do not generalize across hardware and control strategies: Vidur TPOT error reaches 289\% on Qwen3-32B, and AIConfigurator E2E error ranges from 62--97\% (vLLM) and 80--107\% TPOT (SGLang). Vidur isolates the two failure modes \sysname avoids: even with manually registered H200 kernel profiles (Llama-3.1-70B and Qwen3-32B), errors span 39.9--172.2\% E2E and 16.9--288.6\% TPOT---a wide, unstable range---indicating that control-plane re-implementation alone is a major drift source; when profile portability also breaks---for Qwen2.5-72B, calibrated on an older vLLM version without H200 data---errors rise further to 179.5\% E2E and up to 250.3\% TPOT.}{Among the H200 cells with available baseline results, Vidur reaches 172.2\% E2E PE and 288.6\% TPOT PE on Qwen3-32B, while AIConfigurator's vLLM E2E PE spans 62.1--97.0\% and its SGLang TPOT PE spans 69.5--107.2\%. These results characterize observed portability and coverage under our compatibility setup, with each simulator evaluated on its available configuration set.}

\yongchao{[ranking validity] Please report the number of ranked points and whether configurations are pooled across models/platforms. A pooled coefficient may be dominated by obvious between-model scale differences; search utility needs within-model/platform correlation, top-$k$ recall, or selected-configuration regret.}

\subsection{Component Validation} \label{sec:component_ablation}

\ycadd{
\subsubsection{Stage-Wise vs.\ Monolithic Modeling} \label{sec:stage_ablation}
We isolate the effect of the stage decomposition using a separate dense-model trace bank collected on H200. We compare two execution predictors trained from the same traces: a stage-wise model that predicts pre-forward, forward, and post-forward targets separately, and a monolithic model that predicts whole-iteration latency directly. Both models use the same configuration-level three-fold split and hyperparameter optimization strategy.
}

\begin{figure}[t]
  \centering
  \includegraphics[width=\columnwidth]{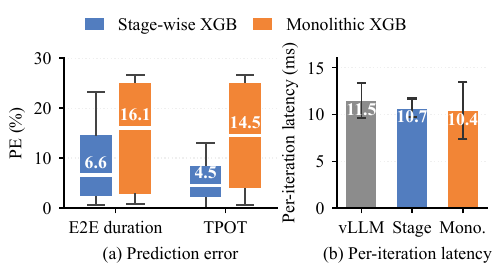}
  \caption{Stage-wise versus monolithic execution modeling on held-out Qwen3-14B/TP1. (a) E2E and TPOT PE across six configurations; (b) mean per-iteration latency, with error bars denoting standard deviation across configurations.}
  \label{fig:stagewise_ablation}
\end{figure}

\ycadd{
We evaluate both alternatives on Qwen3-14B at TP=1, with both the model and serving configurations held out from training. As shown in Figure~\ref{fig:stagewise_ablation}, stage-wise modeling reduces the median E2E PE by $2.4\times$ and the median TPOT PE by $3.2\times$ compared with the monolithic predictor, while both predict mean per-iteration latency within 10\% of vLLM. These results show that retaining stage-level execution structure improves end-to-end prediction under model and configuration shifts.
}

\subsubsection{Control-Plane Fidelity under a Policy Change} \label{sec:control_fidelity}

\ycrep{
We evaluate whether \sysname continues to reproduce framework control behavior after a preemption-policy change. We replace vLLM's original rule, which preempts the tail request in FCFS order, with one that preempts the request occupying the most KV-cache space. We then run the same \sysname implementation with either the modified policy or the original FCFS policy through the existing control-plane hooks, and compare both against Vidur's re-implemented vLLM scheduler.}{We replace vLLM's FCFS tail-preemption rule with max-KV preemption and use physical execution of the modified vLLM as reference. We compare \sysname using the modified native policy, \sysname retaining FCFS, and Vidur under the same request trace.}

\yongchao{[must-fix] This is the paper's strongest distinctive claim, yet it is validated only by KV-cache footprint and iteration count. Add decision-level matching against physical execution: per-iteration batch composition, admission/preemption events, and completion order. Compare both by scheduler iteration and aligned wall/virtual time, because iteration indexing can hide timing drift, and repeat under multiple bursty arrival traces. If decisions diverge while aggregates match, explain why rather than letting coincidence support the claim.}

\yongchao{[missing comparison definition] The results later introduce a `reference' KV-cache trajectory, but this paragraph names only two \sysname variants and Vidur. Please explicitly describe the physical modified-vLLM reference run and how scheduler iterations or timestamps are aligned across systems.}

\begin{figure}[t]
  \centering
  \includegraphics[width=\columnwidth]{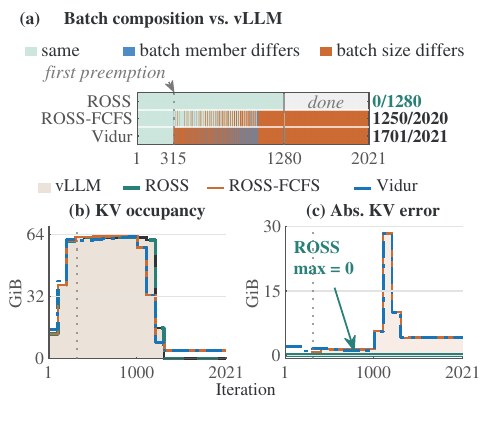}
  \caption{\ycrep{Control-plane fidelity under the modified preemption policy. (left) Mean KV-cache occupancy. (right) Non-empty scheduler iterations to complete the workload.}{Native-control fidelity under modified max-KV preemption. (a) Batch-composition differences from vLLM; (b) KV-cache occupancy; (c) absolute occupancy error. Fractions in (a) denote differing/total non-empty scheduler iterations.}}
  \yongchao{[figure clarity] The paper's strongest claim---`\sysname closely follows the reference KV-cache footprint'---depends on this figure, but in the left panel the \sysname line is visually indistinguishable from the reference (vLLM modified) shaded area. A skeptical reviewer will say: `I can't see the \sysname line, so I can't verify the claim.' Either (a) plot \sysname as a bold colored line \emph{over} the reference shaded region so the overlap is visible, or (b) add a third panel showing the per-iteration \emph{difference} between \sysname and reference (should be near zero).}
  \label{fig:control_fidelity}
\end{figure}

Figure~\ref{fig:control_fidelity} compares decision- and state-level fidelity. \sysname with the modified native policy matches all 1,280 reference batch compositions and all 49 preemption victims, with zero maximum KV-cache occupancy error.
In contrast, \sysname-FCFS and Vidur differ from the reference in 1,250/2,020 and 1,701/2,021 scheduler iterations, respectively.
Native scheduler reuse thus enables \sysname to inherit the modified preemption policy.

\subsubsection{Closed-Loop Traffic Fidelity}
\label{sec:traffic_ablation}

\ycadd{
We evaluate whether \sysname preserves closed-loop traffic behavior when request admission depends on simulated completion. Using the settings in Table~\ref{tab:eval_overview}, we compare vLLM and \sysname under the same frozen request sequence. Besides admission and completion timestamp errors, we measure the deviation in the number of in-flight requests. Let $A(t)$ and $C(t)$ denote cumulative admissions and completions, respectively, and $I(t)=A(t)-C(t)$. We define the \emph{In-flight Count Error} as
\begin{equation}
E_{\mathrm{IF}}
=
\frac{1}{KT}
\int_0^T
\left|
I_{\mathrm{\sysname}}(t)-I_{\mathrm{phys}}(t)
\right|\,dt
\times 100\%,
\end{equation}
where $K$ denotes the concurrency limit and $T$ denotes the evaluation horizon over which the integral is computed.}

\begin{figure}[t]
  \centering
  \includegraphics[width=\columnwidth]{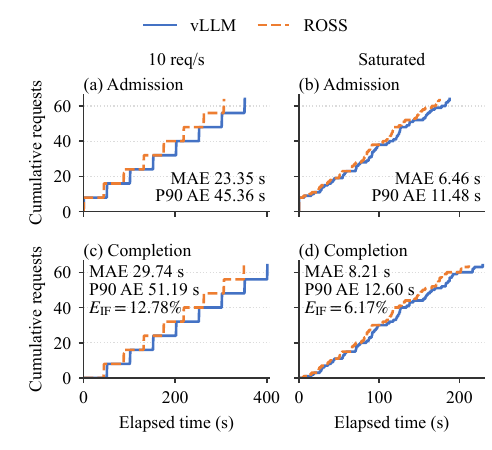}
  \caption{\ycadd{Closed-loop traffic fidelity under 10 req/s and saturated, concurrency-limited workloads. Panels compare cumulative admissions and completions between physical vLLM and \sysname; annotations report mean absolute error (MAE), 90th-percentile absolute error (P90 AE), and $E_{\mathrm{IF}}$.}}
  \label{fig:traffic_fidelity}
\end{figure}

\ycadd{
Figure~\ref{fig:traffic_fidelity} shows that \sysname follows the physical admission and completion trajectories in both regimes, with $E_{\mathrm{IF}}=12.78\%$ at 10 req/s and $6.17\%$ under saturation. Timestamp deviations also include execution-latency prediction error, because a shifted completion correspondingly shifts the admission it releases.
}

\subsubsection{Decode-Window Coalescing} \label{sec:coalescing_ablation}

\ycadd{
We evaluate whether decode-window coalescing reduces repeated control and prediction overhead while preserving predicted E2E duration. We replay the same frozen workload with coalescing enabled and disabled and all serving parameters fixed (Table~\ref{tab:eval_overview}).
}

\begin{figure}[t]
  \centering
  \includegraphics[width=\columnwidth]{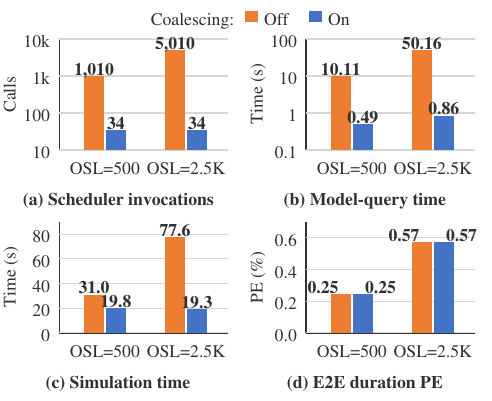}
  \caption{Decode-window coalescing for two OSLs.}
  \label{fig:coalescing_ablation}
\end{figure}

\ycadd{
Figure~\ref{fig:coalescing_ablation} shows that coalescing reduces scheduler invocations from 1,010 to 34 at OSL 0.5K and from 5,010 to 34 at OSL 2.5K. Correspondingly, total simulation time decreases from 31.0\,s to 19.8\,s ($1.57\times$) and from 77.6\,s to 19.3\,s ($4.02\times$), while model-query time exhibits the same trend. Coalescing leaves the E2E duration unchanged in both workloads. Therefore, coalescing amortizes repeated scheduler and predictor invocation overhead while preserving E2E prediction in both workloads; its runtime benefit increases with output length.
}




\subsection{Long-Context Prefill Simulation Performance} \label{sec:prefill}

Long-context prefill is a distinct simulation regime: TTFT depends on prompt-processing cost, while prefix reuse changes how much of the prompt requires computation\ycdel{when each request reaches the native scheduler.
We therefore evaluate whether \sysname remains accurate as context length and prefix reuse vary jointly}.
\ycdel{
\para{Setup.}
We run Qwen3-235B-A22B on B200 with SGLang and DP:PP:TP=1:1:8. To isolate long-context prefill fidelity from queueing and decode interference, we issue 128 requests at 1 request/s, cap the maximum batch size at 16, and fix OSL at 50 tokens (queueing and batch-size robustness are evaluated separately in \S\ref{sec:generalization}).
}
\ycrep{
We use SGLang's native generated-shared-prefix (GSP) workload generator from \texttt{bench\_serving}~\cite{sglang_bench_serving}. Requests are organized into 16 groups of eight prompts, each consisting of a group-shared prefix and a request-specific suffix. We vary their relative lengths to sweep
the aggregate prefix-cache hit rate from 0\% to 80\% in 20-percentage-point increments, while sweeping ISL from 2.5K to 20K tokens.
}{
To isolate long-context prefill behavior from queueing and decode interference, we use low-rate SGLang generated-shared-prefix (GSP) traffic with short outputs and a limited batch size; requests within each group share a common prefix and retain request-specific suffixes. We jointly vary context length and configured prefix-cache hit rate (Table~\ref{tab:eval_overview}).
}

\begin{figure}[t]
  \centering
  \includegraphics[width=\columnwidth]{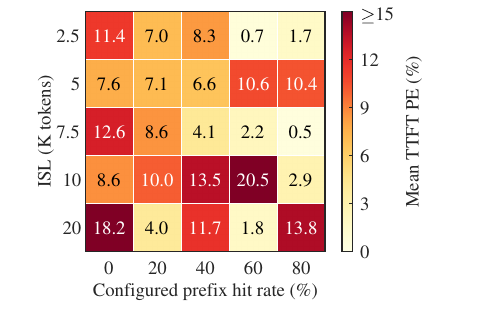}
  \yongchao{[figure placement] This long-context figure is forced into the middle of the control-fidelity subsection, two subsections after its result paragraph. Move the entire figure block to §5.3 immediately after the TTFT result so the visual remains connected to the claim it supports.}
  \caption{Mean TTFT PE across long-context prefill configurations \ycrep{on SGLang (B200, Qwen3-235B-A22B, DP:PP:TP=1:1:8, OSL=50)}{varying context length and prefix-cache reuse}.}
  \yongchao{[figure clarity] The heatmap encodes values purely by color, making precise comparison across cells difficult (e.g., is a cell 8\% or 12\%?). Adding small numerical annotations inside each cell would let the reader read exact values without relying on the color bar. This is standard practice for small heatmaps (5$\times$5) and greatly improves self-sufficiency of the figure.}
  \label{fig:prefix_ttft}
\end{figure}

\para{Results.}
Figure~\ref{fig:prefix_ttft} reports all 25 combinations. The median PE in mean TTFT is 8.3\%, with a p90 of 13.7\%; only 2 of 25 settings exceed 15\%. \ycrep{The errors show no monotonic trend with either context length or
prefix-cache hit rate, indicating that prediction error does not systematically grow over the evaluated long-context range. Therefore, prediction accuracy remains stable across an $8\times$ increase in context length and widely varying degrees of prefix reuse.}{The error in mean TTFT shows no systematic degradation with either context length or prefix-cache hit rate over the evaluated range.
}
\yongchao{[metric coverage] Long-context fidelity is evaluated only through error in mean TTFT, although deployment decisions are usually governed by tail TTFT. With only 128 requests per setting, report per-request distributional accuracy and p95 TTFT error, particularly at 20K context and high prefix reuse; add repeated workloads before making stronger extreme-tail claims.}

\subsection{Generalization and Robustness} \label{sec:generalization}

\begin{figure*}[t]
  \centering
  \includegraphics[width=\textwidth]{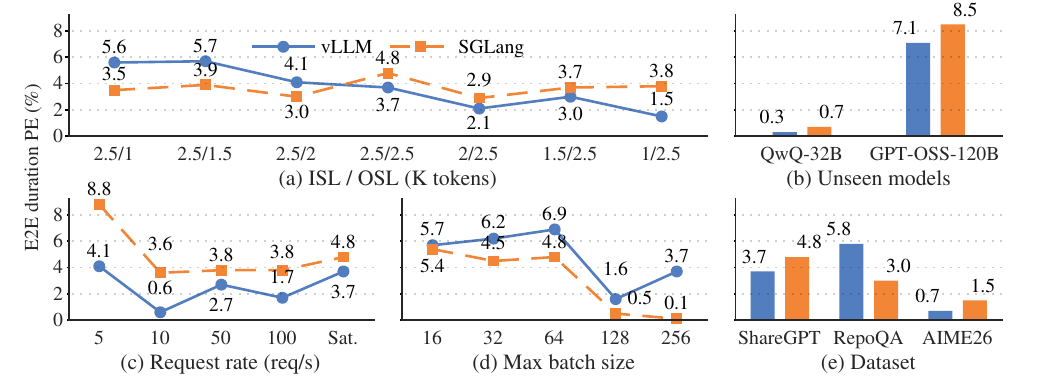}
  \caption{E2E duration PE across five \ycrep{generalization and robustness dimensions}{robustness dimensions}: (a) ISL/OSL, (b) calibration-held-out models, (c) request rate, (d) maximum batch size, and (e) dataset source. In (d), SGLang points at batch sizes 128 and 256 use the TokenizerManager patch.}
  \label{fig:generalization_summary}
\end{figure*}

\yongchao{[generality] Four non-model sweeps---ISL/OSL, request rate, batch size, and dataset---share the Llama-3.1-70B/B200 default, while the two held-out models are tested at only one operating point. Replicate at least the ISL/OSL and request-rate sweeps for one MoE model, so the paper shows that robustness dimensions compose rather than hold only in isolation.}
Figure~\ref{fig:generalization_summary} summarizes generalization across \emph{workload patterns (ISL/OSL)}, \emph{unseen models}, \emph{request rates}, \emph{batch sizes}, and \emph{dataset distributions}. \ycrep{ These experiments test whether the data-plane regressor generalizes across unseen workload and model conditions, and whether native control invocation preserves serving behavior as arrivals and batch formation change.
}{ Unless otherwise noted, each experiment varies one dimension from the default configuration in Appendix Table~\ref{tab:eval_setup_full}. These sweeps evaluate prediction robustness across workload, model, and serving conditions. }

\ycadd{As shown in Figure~\ref{fig:generalization_summary}, E2E PE remains 1.5--5.7\% on vLLM and 2.9--4.8\% on SGLang across ISL/OSL mixes; on the calibration-held-out QwQ-32B and GPT-OSS-120B models it remains below 1\% and 9\%, respectively. Across request-rate sweeps, maximum batch sizes 16--256, and three dataset sources, E2E PE does not exceed 8.8\% and is single-digit in all evaluated settings.}

\subsection{Simulation Throughput and Multi-Instance Scaling} \label{sec:scalability}

\ycrep{
This subsection validates the remaining asymmetry claims directly: O(1) execution lookup yields over $11\times$ speedup over real hardware, the single-process simulation loop keeps runtime flat as host CPUs become contended, and virtual-time semaphores preserve correct back-pressure under concurrency caps. Together, these properties allow a CPU server to evaluate many deployment candidates concurrently without physical GPU execution.
}{
This subsection evaluates the runtime efficiency and scaling behavior of \sysname. We compare its wall-clock runtime with real execution and measure it independently under reduced CPU resources and when multiple simulation instances execute concurrently. Runtime experiments use a dual-socket Intel Xeon Platinum 8468 server (96 physical cores), 2\,TB of DRAM, and 2 NUMA nodes. For the core sweep, we vary the CPU-affinity mask from 2 to 32 physical cores while keeping one simulation worker and one internal thread.}
\yongchao{[missing experiment] No result below varies the concurrency cap or compares back-pressure behavior. Add a semaphore-limit sweep against real execution using admission times, queue length, throughput, and tail latency, or remove the claim that correctness is validated here.}

\para{Single-instance speedup.}
\ycrep{
\sysnames computation--control asymmetry predicts that simulation time should be dominated by the CPU-only control plane, making it much faster than real execution even without GPU acceleration. Table~\ref{tab:speedup} confirms: under 32 cores, \sysname completes a 512-prompt workload in 20.5\,s versus 243.2\,s real execution ($11.9\times$ speedup) and 166.7\,s CPU-only time ($8.1\times$), because \sysname skips model-weight loading and GPU kernel dispatch entirely.
}{Table~\ref{tab:speedup} compares \sysnames wall-clock simulation time with real vLLM execution. With 32 pinned cores, the complete \sysname process finishes the 512-prompt workload in 21.4\,s, versus 243.2\,s
for real execution, corresponding to an $11.4\times$ speedup.}

\para{Scaling under CPU contention.}
\ycrep{
Revati does not currently provide a public implementation. We therefore characterize time-warp emulation using an optimistic analytical bound: the CPU-only component of real serving execution provides a lower bound on time-warp runtime and thus an upper bound on its speedup.
}{
With no public Revati implementation available at the time of our evaluation, we report the E2E--GPU residual, $\widetilde{T}_{\mathrm{res}}=T_{\mathrm{real}}-T_{\mathrm{GPU}}$, as a heuristic reference for GPU-free native-runtime emulation, where $T_{\mathrm{GPU}}$ is accumulated measured GPU execution time.
This heuristic ignores CPU--GPU overlap and assumes unchanged non-GPU runtime costs.
}
\yongchao{[baseline validity] CPU and GPU work overlap, so `real E2E minus GPU time' is not CPU-only execution time and is not a proven lower bound on time-warp runtime. Derive the bound from a measured critical-path decomposition, or present it only as a heuristic estimate rather than an upper bound.}

\ycdel{Multi-instance simulation on a shared CPU server reduces per-instance core allocation, stressing control-plane scalability. }As cores drop from 32 to 2, \ycdel{real vLLM's CPU time grows from 166.7\,s to 1077.6\,s while GPU time stays roughly constant (the slight non-monotonicity at 2 cores reflects CPU-starved batch formation causing the GPU to occasionally idle between steps); \sysnames runtime stays flat at 20.5\,s, widening the speedup to $52.6\times$ vs.\ real and $45.9\times$ vs.\ CPU time.}
\yongchao{[ numerical consistency] This prose conflicts with Table~\ref{tab:speedup}: 1,077.6 seconds is Real E2E, whereas the two-core CPU Time is 940.2 seconds; GPU time also rises monotonically from 76.5 to 137.4 seconds (about 80\%), and no two-core non-monotonicity is shown. Correct the labels and explain why measured GPU time rises under CPU contention.}\ycrep{This is exactly the multi-instance, resource-contended regime where large-scale configuration search operates.}{real vLLM E2E duration increases from 243.2\,s to 1,077.6\,s, while the heuristic $\widetilde{T}_{\mathrm{res}}$ increases from 166.7\,s to 940.2\,s. Accumulated GPU execution time also increases from 76.5\,s to 137.4\,s, indicating that GPU-side execution is not invariant to host contention. \sysnames measured runtime remains between 21.4\,s and 22.1\,s across the core-count sweep, yielding a 48.7$\times$ speedup against real execution at two cores.
The heuristic reference is 7.8--42.5$\times$ \sysnames measured runtime across the sweep.}

\para{Multi-instance scaling.}
\sysname runs each simulation in a single-process event loop without framework-level multiprocessing, inter-process communication, or physical device state. Independent simulations can therefore execute concurrently across CPU cores. \ycrep{Using the same per-task configuration as Table~\ref{tab:speedup}, parallelizing 64 simulations across 32 processes achieves a $20.0\times$ speedup, reducing the sequential baseline from 1,446.4\,s to 72.4\,s.}{Using the same workload and serving configuration as the single-instance experiment, running 64 simulations with 32 concurrent processes reduces wall-clock time from 1,370.5 to 72.4\,s, an 18.9$\times$ speedup.}

\yongchao{[numerical consistency] The claimed `same per-task configuration' is inconsistent with Table~\ref{tab:speedup}: $64\times20.5$ seconds equals 1,312 seconds, whereas 1,446.4 seconds implies 22.6 seconds per run. Moreover, the displayed real/runtime values give $(64\times243.2)/72.4\approx215\times$, not 238$\times$. Recompute all speedups from one consistent baseline and state the included overheads.}
\yongchao{[implementation] The multi-instance result lacks the host configuration needed to interpret it. Report physical/logical core count, CPU model, memory, process placement, and utilization, then show measured end-to-end time for the 228-candidate search rather than extrapolating practical throughput from an underspecified 64-task run.}
\ycdel{Relative to executing the same 64 workloads sequentially on real hardware, the 72.4\,s parallel-simulation makespan yields an aggregate throughput gain of ${\sim}215\times$.
For reference, dividing real E2E duration by the heuristic E2E--GPU residual gives $1.15$--$1.50\times$ across the core-count sweep; we do not interpret this ratio as a formal bound on time-warp speedup.}

\yongchao{[scaling evidence] Table~\ref{tab:speedup} supplies one \sysname runtime spanning all five core-count columns, yet the text concludes runtime is flat from 32 to 2 cores. Run \sysname under explicit CPU pinning or cgroup limits at every core count and report independently measured runtimes with variance.}

\begin{table}[t]
\centering
\caption{\ycrep{CPU budget vs.\ \sysname speedup (B200, Llama-3.1-70B, 1:1:8, batch=256, ISL/OSL=2500/2500). CPU Time = real E2E minus GPU time, an optimistic lower bound on TW runtime.}{Runtime under CPU constraints. $\widetilde T_{\mathrm{res}}$ denotes the E2E--GPU residual defined in the text.}}
\label{tab:speedup}
\small
\begin{tabular*}{\columnwidth}{@{\extracolsep{\fill}}lrrrrr@{}}
\toprule
Cores & 32 & 16 & 8 & 4 & 2 \\
\midrule
Real E2E (s)      & 243.2 & 246.9 & 307.9 & 423.2 & 1077.6 \\
Accum. GPU (s)      & 76.5  & 77.0  & 102.8 & 117.1 & 137.4  \\
$\widetilde{T}_{\mathrm{res}}$ (s)      & 166.7 & 169.9 & 205.2 & 306.1 & 940.2 \\
\midrule
\sysname (s)              & $21.4$ & $21.7$ & $22.1$ & $21.8$ & $22.1$ \\
vs.\ real ($\times$)      & 11.4 & 11.4 & 13.9 & 19.4 & 48.7 \\
vs.\ $\widetilde{T}_{\mathrm{res}}$ ($\times$)  & 7.8  & 7.8  & 9.3 & 14.0 & 42.5 \\
\bottomrule
\end{tabular*}
\end{table}

\subsection{Configuration-Search Case Study} \label{sec:use_case}
\yongchao{[evaluation ordering] The two application case studies appear before the experiment that establishes \sysname's own simulation throughput and scaling. Move §5.8 before the case studies, so the paper first validates accuracy, component fidelity, and efficiency, then uses those established properties to support search and diagnosis.}
A key operational use case is guiding deployment configuration search under joint efficiency--user-experience trade-off constraints~\cite{stojkovic2025dynamollm}, which often requires evaluating hundreds of candidate configurations before a production launch. We demonstrate this with a \emph{Pareto-frontier} search over both co-located and disaggregated deployment modes.

\begin{figure}[t]
  \centering
  \includegraphics[width=\columnwidth]{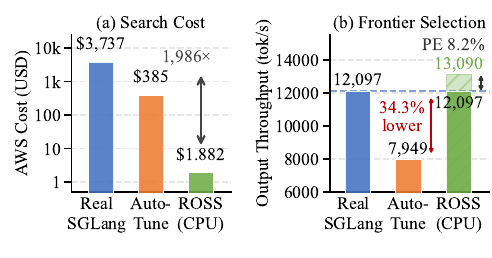}
  \caption{Configuration search cost and selected throughput for Qwen3-32B under an eight-B200 budget.
  \textbf{(a)} AWS cost estimated from on-demand rates. \textbf{(b)} \ycrep{Best output throughput found; the hatched bar shows \sysnames \emph{predicted} throughput (10{,}589 tok/s, PE\,7.3\% above real)}{Measured throughput of selected configurations; \sysnames prediction is hatched}.}
  \label{fig:search_cost}
\end{figure}

\para{Search setup.}
We search 228 configurations spanning parallelism, memory fraction, and chunked-prefill size. We filter candidates using a 100\,s SLA on P99 request-level E2E latency and construct a Pareto frontier within each deployment mode over \emph{tokens/s/user} and \emph{tokens/s/GPU}. We validate all nine predicted frontier configurations on real SGLang and compare against an SGLang adaptation of Auto-Tune~\cite{vllm_autotune}, which searches nine co-located TP$=8$ configurations under the same SLA.
Figure~\ref{fig:search_cost} summarizes the results; Appendix~\ref{app:pareto} details the search-space construction and Auto-Tune procedure.

\yongchao{[use-case validity] The setup promises a two-objective Pareto search, but the results show only one maximum-throughput point and compare it with a five-probe scalar Auto-Tune run. Show predicted versus real frontiers, frontier recall/hypervolume, the deployment-selection policy, and selected-configuration regret or top-candidate margins that demonstrate the 7.3\% prediction bias did not change the decision. Give Auto-Tune the same objective and hardware-evaluation budget; otherwise recast this as scalar throughput optimization and avoid attributing its miss solely to the algorithm.}
\yongchao{[cost accounting] The \sysname cost counts only the CPU sweep and excludes required hardware microbenchmarks and serving-trace collection. Report both cold-start and amortized cost, include all GPU-hours, and state how many searches share one profile before claiming a 1,258$\times$ reduction.}

\para{Search cost and quality.}
\sysname simulates all 228 candidates in ${\sim}$70 minutes on a CPU server using 16 concurrent simulations, versus ${\sim}$32.8 B200 node-hours (262.4 GPU-hours) for exhaustive evaluation and ${\sim}$203 minutes for Auto-Tune. Applying AWS on-demand rates to these runtimes gives estimated costs of \$1.882, \$3,737, and \$385, respectively, yielding a $1{,}986\times$ reduction in \sysnames marginal search cost relative to exhaustive evaluation.\footnote{\url{https://aws.amazon.com/ec2/pricing/on-demand/}; AWS on-demand pricing (us-east-1, accessed September 23, 2026): p6-b200.48xlarge (8$\times$B200), \$113.9328/h; m7i.8xlarge (32 vCPU), \$1.6128/h.} The CPU estimate assumes comparable runtime on the priced instance; \sysnames cost excludes one-time profiling/predictor construction and post-search hardware validation.

Among the nine hardware-validated frontier configurations, the one with the highest predicted aggregate output throughput achieves 12,097 tokens/s on real SGLang (8.2\% PE), which is also the highest measured throughput among the validated configurations. Auto-Tune's selected configuration achieves 7,949 tokens/s, 34.3\% lower than the \sysname-selected configuration's throughput.
Thus, the broader simulation-based search identifies a higher-throughput SLA-feasible configuration.

\subsection{Performance Diagnosis Case Study} \label{sec:patching}
\yongchao{[scope justification] \sysname does not model detokenization and response dispatch, but the paper does not explain whether this is a principled boundary or an implementation limitation. Quantify the post-generation share of E2E duration across the evaluated cells, identify the conditions where it becomes material, and state how excluding it changes the scope of the headline fidelity claim.}
A complementary use case for \sysname is diagnosing unexpected discrepancies between simulated and real serving performance. Currently, \sysname does not model the post-generation path (e.g., detokenization and response dispatch). \ycrep{Although this omission does not dominate the aggregate errors in most evaluated settings (\S\ref{sec:e2e}), it can become material at specific operating points.}{This omission can become material at specific operating points.} \ycrep{When the post-generation latency spikes at specific operating points, \sysnames stage-level decomposition exposes the gap as an anomaly concentrated outside the modeled stages, isolating a real framework bottleneck rather than a simulation modeling error.}{Such simulation--measurement discrepancies can therefore provide a useful signal for identifying performance bottlenecks outside the modeled execution path.}
\yongchao{[causal precision] The decomposition reveals an unexplained residual outside the modeled stages; only the subsequent instrumentation attributes it to TokenizerManager. Please separate anomaly detection from root-cause identification instead of claiming that the decomposition alone isolates a real framework bottleneck.}

\para{Observation.}
\ycrep{During fidelity evaluation on SGLang, \sysnames stage-level decomposition flagged a consistent discrepancy: simulated E2E duration was lower than real measurements at batch size 256, concentrated in the post-generation path after decode completes but before client delivery.}{Against vanilla SGLang, \sysname underpredicts E2E duration at maximum batch sizes 256 and 512. As post-generation processing is outside \sysnames modeled path, we investigate whether this discrepancy exposes a bottleneck.}

\para{Root cause.}
Instrumentation confirmed a \emph{batch-boundary burst}. \ycrep{when requests complete near-simultaneously, the DetokenizerManager floods the TokenizerManager's single event loop (message reception, state updates, HTTP streaming), transiently saturating it and delaying admission of the next prefill batch.}{Near-simultaneous request completions cause the DetokenizerManager to flood the TokenizerManager’s single event loop. The loop handles message reception, state updates, and HTTP streaming; its temporary saturation delays admission of the next prefill batch.}

\para{Patch and outcome.}
A structural fix to \texttt{TokenizerManager} separates the detokenizer receive path from waiter notification, reducing contention from completion bursts.
Table~\ref{tab:patching} shows workload E2E-duration reductions of 35.8\% and 32.1\% at batch sizes 256 and 512, respectively, with smaller changes at lower batch sizes.

\yongchao{[missing mechanism evidence] The displayed table contains only total E2E duration, so the stage-level localization is not visible. Add observed-versus-predicted pre-forward, forward, post-forward, and post-generation components before/after the patch, including the residual that directed attention to TokenizerManager.}

\begin{table}[t]
\centering
\caption{SGLang E2E duration before/after TokenizerManager patch across maximum batch sizes (Qwen3-32B, 8×B200).}
\label{tab:patching}
\small
\begin{tabular*}{\columnwidth}{@{\extracolsep{\fill}}lrrrrr@{}}
\toprule
Batch size & 32 & 64 & 128 & 256 & 512 \\
\midrule
Before (s) & 538.3 & 215.6 & 150.5 & 139.7 & 130.3 \\
After (s)  & 530.2 & 214.3 & 139.8 & \textbf{89.7} & \textbf{88.5} \\
Improv.\ (\%) & 1.5 & 0.6 & 7.1 & \textbf{35.8} & \textbf{32.1} \\
\bottomrule
\end{tabular*}
\end{table}

\section{Discussion and Future Work}
\label{sec:discussion}

\ycadd{
\para{Post-generation processing.}
\sysname currently does not model post-generation processing. \S\ref{sec:patching} shows that this path can become material under high completion concurrency.}

\ycadd{
\para{Control-plane coverage.}
Our evaluation exercises state-dependent scheduling and KV-cache management through one vLLM preemption-policy change on one model; broader policy and framework-version changes remain untested.
PD serving abstracts KV transfer as a zero-delay logical handoff; accuracy when transfer latency or network contention dominates is not evaluated.}

\ycadd{
\para{Hardware portability.}
Adapting to a new hardware platform requires platform-specific profiling to obtain the descriptors used by the regressors. Evaluation covers H200/B200; broader cross-platform transfer remains future work.}

\ycadd{
\para{Beyond transformer architectures.}
Our evaluation covers transformer-based dense and MoE models. Architectures with substantially different execution structure, such as state-space models (SSMs), are not evaluated and may require additional modeling features or stage definitions.}

\section{Conclusion}
\yongchao{[structural] The conclusion contains no limitations paragraph. For a systems paper at this venue, reviewers expect at least a brief acknowledgment of scope: (1) ROSS does not model post-generation paths (the SGLang diagnosis shows this can matter), (2) control-plane fidelity is validated only for one policy change on one model, (3) the stage-wise regressor requires hardware-specific profiling for each new platform, (4) generalization to non-transformer architectures (e.g., SSMs) is untested. A 4-sentence limitations paragraph preemptively addresses these and signals intellectual honesty.}

\ycrep{
We presented \sysname, an inference simulator that resolves the speed--accuracy trade-off in LLM serving simulation by exploiting the computation--control asymmetry: a stage-wise data plane predicts iteration latency using sparse serving traces and lightweight hardware descriptors, while module-level control-plane emulation preserves framework-native control logic over simulated state (\(\sim\)8.2K LOC, 3 hook points per framework) without full-runtime replay.}{We presented \sysname, a dual-plane LLM serving simulator combining stage-wise execution prediction with framework-native control.}
Across \ycrep{H200/B200}{platforms}, \ycdel{seven} models, and \ycrep{evaluated vLLM/SGLang serving paths}{serving frameworks}, \sysname achieves below 6\% median E2E/TPOT error and over $11\times$ simulation speedup over real execution\ycdel{, while following framework control-policy changes through native module reuse}. A 228-candidate search completes in 70 minutes, and simulation-guided diagnosis of an SGLang bottleneck leads to a 35.8\% workload E2E-duration reduction.
\ycadd{
Broader coverage of serving systems and model architectures remains future work.
}

\clearpage

\bibliographystyle{ACM-Reference-Format}
\bibliography{content/ref}


\clearpage
\appendix

\section{Performance-Model Feature Specification}
\label{app:features}

Table~\ref{tab:feature_spec} specifies the feature schema for the stage-wise
performance predictors in the data plane (\S\ref{sec:4_2}). Let $d_h$ denote the attention-head dimension from the model configuration when available, and $d/n_h$ otherwise. Let $\mathcal{B}$ denote the scheduled batch, $s_i$ the current context length of request $i$, and $C=\sum_{i\in\mathcal{B}}s_i$.  For prefill request $j$, $q_j$ is the number of tokens scheduled in the current iteration and $k_j$ the context length attended by those tokens. We use $\ell(x)=\log_{10}(x)$ for $x>0$ and $\ell(x)=0$ otherwise.
In the scope column, $\mathsf{V}$ denotes the vLLM stage predictors and $\mathsf{S}_{P}$ and $\mathsf{S}_{D}$ the SGLang prefill and decode stage predictors.  A scope covers pre-forward, forward, and post-forward unless stated otherwise.  Numeric features are not standardized; counts, dimensions, and physical descriptors use their native units unless an
explicit logarithm, ratio, or Boolean encoding is shown.  Derived operation-, memory-, occupancy-, and communication-scale quantities are auxiliary descriptors supplied to the learned regressors, not standalone analytical latency estimates.

\sysname trains separate predictor families by framework and latency target;
for SGLang, these families are further specialized for prefill and decode.
Model architecture and platform are predictor inputs rather than family
identifiers, allowing each family to span the profiled models and platforms.
Each predictor consumes the feature families available at its boundary;
batch-dependent values are constructed from native scheduler output after
batch formation and before latency prediction.

\para{Parallelism and model notation.}
We use DP, PP, TP, and EP to denote data, pipeline, tensor, and expert parallelism, respectively. A DP:PP:TP tuple specifies the corresponding parallelism degrees. For PD deployments, $A@B$ denotes prefill tuple $A$ and decode tuple $B$.
FFN and GQA denote feed-forward network and grouped-query attention, respectively.

\para{Platform-descriptor profiling.}
The platform descriptors in Table~\ref{tab:feature_spec} are collected using general matrix multiplication (GEMM), attention, and NVIDIA Collective Communications Library (NCCL) microbenchmarks. On the evaluated 8-GPU platforms, collecting these descriptors requires approximately 900 measurements and 3--4 wall-clock hours (${\sim}$30 GPU-hours). These descriptors are workload-independent and therefore need not be recollected when only workload conditions change.

\section{Configuration-Search Details and Results}
\label{app:pareto}

\begin{figure}[t]
  \centering
  \includegraphics[width=\columnwidth]{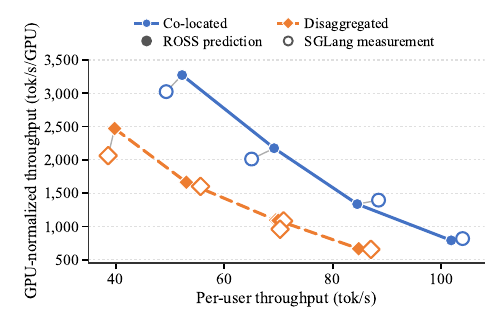}
  \caption{Predicted Pareto frontiers for co-located and disaggregated deployments, with paired SGLang measurements for all nine hardware-validated configurations.}
  \label{fig:pareto_result}
\end{figure}

The configuration-search case study (\S\ref{sec:use_case}) constructs separate Pareto frontiers for co-located and disaggregated deployments over \emph{tokens/s/user} and \emph{tokens/s/GPU} under an eight-GPU B200 budget. We validate all nine predicted frontier configurations on real SGLang; Figure~\ref{fig:pareto_result} shows their paired predictions and measurements.

\para{Search space.}
The 228 candidates comprise 42 co-located configurations (7 DP$\times$TP parallelisms $\times$ 3 memory fractions $\times$ 2 chunked-prefill sizes) and 186 disaggregated configurations (31 prefill--decode TP/DP combinations $\times$ 3 memory fractions $\times$ 2 chunked-prefill sizes). We use memory fractions $\{0.8,0.9,0.95\}$ and chunked-prefill sizes $\{8192,32768\}$ tokens.

\para{Search runtime.}
The 70-minute \sysname search was measured on the CPU server described in \S\ref{sec:scalability}, using 16 concurrent single-process simulations.

\para{Auto-Tune baseline.}
Our SGLang adaptation of Auto-Tune searches nine co-located TP$=8$ configurations by jointly tuning scheduler concurrency and chunked-prefill size. We use the same 100\,s SLA on P99 request-level E2E latency for \sysname and Auto-Tune. Each configuration is first evaluated under saturated load; if it violates the SLA, Auto-Tune reduces the offered request rate until the constraint is satisfied. Among SLA-feasible settings, Auto-Tune selects the configuration with the highest request throughput.

\section{Native-Control Integration}
\label{app:native_control}

Table~\ref{tab:implementation} summarizes the reused framework modules and the \sysname implementation footprint.
Table~\ref{tab:control_state} clarifies the logical state scope maintained across native-control invocation and system-level simulation.

\section{Common Serving Settings}
\label{app:common_settings}

Unless otherwise noted, experiments in \S\ref{sec:evaluation} use a memory fraction of 0.9 and a chunked-prefill size of 8192 tokens. Memory fraction is set through each backend's native knob. In vLLM, \texttt{gpu\_memory\_utilization} bounds total model-executor memory, with profiled non-KV usage (including activations) deducted before sizing the KV cache. In SGLang, \texttt{mem\_fraction\_static} directly budgets model weights and the KV-cache pool, leaving the remainder for activations and CUDA-graph buffers.

\begin{table*}[t]
  \centering
  \caption{Feature families, representations, sources, and scope of the stage-wise latency predictors.}
  \label{tab:feature_spec}
  \small
  \begin{tabular*}{\textwidth}{
    @{\extracolsep{\fill}}
    >{\raggedright\arraybackslash}p{.14\textwidth}
    >{\raggedright\arraybackslash}p{.47\textwidth}
    >{\raggedright\arraybackslash}p{.17\textwidth}
    >{\raggedright\arraybackslash}p{.10\textwidth}
    @{}}
    \toprule
    \textbf{Feature family} &
    \textbf{Principal representation and encoding} &
    \textbf{Source} &
    \textbf{Scope} \\
    \midrule

    Batch composition &
    Batch size; prefill and decode request counts; and the sum, mean, maximum,
    squared-length sum, and squared-length mean of resident context lengths,
    computed separately for the two phases. &
    Native scheduler output and request state. &
    $\mathsf{V},\mathsf{S}_{P},\mathsf{S}_{D}$ \\

    Scheduled prefill work &
    Number of scheduled prefill requests; sum, mean, maximum, squared sum, and
    squared mean of $q_j$; and the context-sensitive quantities
    $\sum_jq_jk_j$, its per-request mean, and $\ell(\sum_jq_jk_j)$. &
    Scheduler token budget and chunk metadata. &
    $\mathsf{V}$ prefill/mixed;
    $\mathsf{S}_{P}$ \\

    Decode state &
    Number of active decode requests (one decode-step token per request),
    together with the sum, mean, maximum, squared sum, and squared mean of
    their resident context lengths. &
    Native scheduler output and request state. &
    $\mathsf{V}$ decode/mixed;
    $\mathsf{S}_{D}$ \\

    \midrule
    Model geometry &
    Layer count $L$, hidden width $d$, attention- and KV-head counts $n_h,n_{\mathrm{kv}}$, attention-head dimension $d_h$, dense-FFN width $d_{\mathrm{ff}}$, vocabulary size, GQA ratio $n_h/n_{\mathrm{kv}}$, and FFN expansion $d_{\mathrm{ff}}/d$; raw dimensions and dimensionless ratios. & Model configuration. &
    $\mathsf{V},\mathsf{S}_{P},\mathsf{S}_{D}$ \\

    Parallelism and topology &
    TP, PP, and EP degrees; node and GPU counts; and intra-/inter-node
    topology descriptors for TP and EP groups, represented as degrees,
    ratios, and Boolean indicators. &
    Deployment configuration. &
    $\mathsf{V},\mathsf{S}_{P},\mathsf{S}_{D}$ \\

    MoE geometry &
    Expert count $E$, routing fanout $r$, expert width $d_e$, shared-expert
    width $d_s$, and active routed width $r d_e$. &
    Model configuration. &
    MoE variants of
    $\mathsf{V},\mathsf{S}_{P},\mathsf{S}_{D}$ \\

    \midrule
    Workload--model scale &
    Log-transformed total and per-rank projection/FFN operation-scale
    descriptors, together with model-size and activation-volume descriptors
    derived from $C$, model geometry, TP, and PP. &
    Batch, model, and deployment state. &
    $\mathsf{V},\mathsf{S}_{P},\mathsf{S}_{D}$ \\

    Prefill operation scale &
    Log-transformed descriptors proportional to
    $(L/\mathrm{TP})d\sum_jq_jk_j$ for context-dependent attention and
    $(L/\mathrm{TP})(\sum_jq_j)d d_{\mathrm{act}}$ for active FFN work,
    where $d_{\mathrm{act}}$ is the effective active FFN width obtained from
    the model configuration. &
    Scheduled prefill, model, and TP state. &
    $\mathsf{V}$ prefill/mixed;
    $\mathsf{S}_{P}$ \\

    KV and attention working set &
    Log-transformed total and per-rank KV-size descriptors, including a term
    proportional to $Ln_{\mathrm{kv}} d_h C$, and attention activation/weight-volume
    descriptors derived from resident contexts, head geometry, TP, and PP. &
    Context lengths, model geometry, and parallelism. &
    $\mathsf{V},\mathsf{S}_{P},\mathsf{S}_{D}$ \\

    MoE routing and computation &
    Routing sparsity and routing-load proxies derived from active tokens, routing fanout, and expert count, together with router, routed-expert, and shared-expert operation-scale descriptors. &
    Batch, model, and EP state. &
    MoE variants of
    $\mathsf{V},\mathsf{S}_{P},\mathsf{S}_{D}$ \\

    Communication demand &
    Log-transformed total and inter-node TP/EP message-volume descriptors
    and bandwidth--latency service descriptors, derived from active tokens,
    model geometry, deployment topology, and measured collective performance. &
    Batch, model, deployment, and platform descriptors. &
    Distributed variants of
    $\mathsf{V},\mathsf{S}_{P},\mathsf{S}_{D}$ \\

    \midrule
    Platform descriptors &
    Measured GEMM and attention throughput, measured memory bandwidth,
    intra-/inter-node all-reduce and all-to-all bandwidth and latency, and
    theoretical compute and memory-bandwidth peaks.  Values use FLOP/ms,
    GB/s, and ms. &
    One-time GEMM/attention and NCCL microbenchmarks; device specifications. &
    $\mathsf{V},\mathsf{S}_{P},\mathsf{S}_{D}$ \\
    \bottomrule
  \end{tabular*}
\end{table*}

\begin{figure}[t]
  \centering
  \includegraphics[width=\columnwidth]{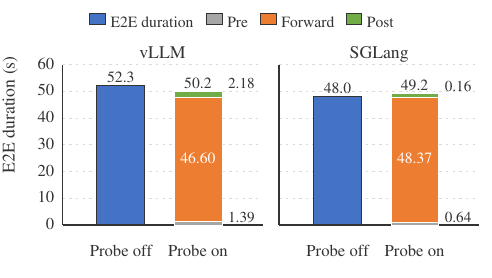}
  \caption{Probe-on/off comparison for framework-boundary instrumentation on vLLM and SGLang (setup in Appendix Table~\ref{tab:eval_setup_full}). Stacked bars decompose probe-on execution into pre-forward, forward, and post-forward timing targets.}
  \label{fig:probe_overhead}
\end{figure}

\begin{table}[t]
\centering
\caption{Control-plane state scope in \sysname.}
\label{tab:control_state}
\begin{tabularx}{\columnwidth}{@{}p{0.25\columnwidth}X@{}}
\toprule
State scope & Included control information \\
\midrule

Request state &
Request identity and serving progress, including prompt and
generated-token progress, request completion status and phase/handoff
information for requests moving across serving stages.
\\[0.4em]

KV/block state &
Logical KV-resource capacity and availability, together with
request-associated allocation state relevant to scheduling,
preemption, completion, and cross-stage serving.
\\[0.4em]

Worker/pool state &
Request placement, worker- or pool-level load information, and
the logical organization of serving workers. PD deployments use
separate prefill and decode pools together with the state needed
to coordinate inter-pool handoff.
\\

\bottomrule
\end{tabularx}
\end{table}

\begin{table}[t]
  \centering
  \caption{\sysname implementation summary. Native control implementations are invoked through \sysname adapters.}
  \label{tab:implementation}
  \footnotesize
  \begin{tabular*}{\columnwidth}{@{\extracolsep{\fill}}llr@{}}
  \toprule
  Component & Source & LOC \\
  \midrule
  Simulation core & \sysname & ${\sim}$1.7K \\
  Scheduler + KV-pool shim & \sysname & ${\sim}$1.1K \\
  Perf.\ profiler & \sysname & ${\sim}$1.1K \\
  Perf.\ regressor + features & \sysname & ${\sim}$0.8K \\
  Predict / bench entry points & \sysname & ${\sim}$1.3K \\
  Pareto optimizer & \sysname & ${\sim}$0.5K \\
  Request dispatcher + HTTP & \sysname & ${\sim}$0.4K \\
  Bench infra \& misc & \sysname & ${\sim}$1.3K \\
  \midrule
  \textit{Reused: Scheduler logic} & vLLM / SGLang & unmod. \\
  \textit{Reused: BlockManager / KV} & vLLM / SGLang & unmod. \\
  \textit{Reused: Load balancer} & vLLM / SGLang & unmod. \\
  \midrule
  Hook points per framework & & 3 \\
  Total \sysname LOC & & ${\sim}$8.2K \\
  \bottomrule
  \end{tabular*}
\end{table}

\begin{table*}[t]
\centering
\caption{
Representative control-relevant changes in vLLM from v0.10.0 to
v0.11.2, the version used in our evaluation~\cite{vllm_github}. The table highlights
upstream changes that alter control state, interfaces, or decision
semantics and would therefore require corresponding updates in a
separately re-implemented simulator.
}
\label{tab:vllm_evolution}
\small
\setlength{\tabcolsep}{4pt}
\renewcommand{\arraystretch}{1.08}

\begin{tabularx}{\textwidth}{@{}lXXX@{}}
\toprule
Version &
Upstream change &
Affected control state / interface &
Simulator-side implication \\
\midrule

v0.10.0 &
Async scheduling (\#19970) &
One-step-ahead scheduling and pending-output state &
Reproduce scheduler--execution ordering and
scheduled-but-not-yet-generated request state \\

v0.10.1 &
Scheduler-side KVConnector output (\#22157) &
Completion notification and finished-request state &
Track the new connector--scheduler callback and corresponding
request-state updates \\

v0.10.2 &
Priority-preemption fix (\#23713) &
Running-request state after KV-pressure eviction &
Update preemption transitions and scheduler running-set semantics \\

v0.11.1 &
Async scheduling with preemption / chunked prefill
(\#26385, \#28787) &
Token history and resumed-request state after preemption &
Reproduce resumed-request and request-history transitions \\

v0.11.1 &
ARC eviction policy for CPU KV-offload cache (\#27039) &
Replacement metadata and cache-victim selection &
Implement the new eviction policy and associated KV-state transitions \\

\bottomrule
\end{tabularx}
\end{table*}

\begin{table*}[t]
  \centering
  \caption{
  Evaluation setup by experiment family. \textsuperscript{D} denotes the default setting for the one-dimensional sweeps in \S\ref{sec:generalization}. Framework versions are given in \S\ref{sec:exp_setup}. Unless otherwise noted, common serving settings follow Appendix~\ref{app:common_settings}.}
  \label{tab:eval_setup_full}

  \small
  \setlength{\tabcolsep}{3pt}
  \renewcommand{\arraystretch}{1.08}

  \begin{tabularx}{\textwidth}{@{}
    >{\raggedright\arraybackslash}p{.105\textwidth}
    >{\raggedright\arraybackslash}p{.17\textwidth}
    >{\raggedright\arraybackslash}p{.06\textwidth}
    >{\raggedright\arraybackslash}p{.05\textwidth}
    >{\raggedright\arraybackslash}X
    >{\raggedright\arraybackslash}X
    @{}}
    \toprule
    Experiment & Model & Backend & GPU & Workload & Serving configuration \\
    \midrule

    \S\ref{sec:modeling_method} Probe sanity
    & Qwen3-32B
    & vLLM / SGLang
    & H200
    & Random; ISL/OSL$=$0.5K/2.5K; saturated; 8 requests
    & DP:PP:TP$=$1:1:1; max batch$=$8; probes on/off
    \\

    \S\ref{sec:e2e} Overall fidelity
    & Seven main-evaluation models
    & vLLM / SGLang
    & H200 / B200
    & ShareGPT; ISL/OSL$=$0.5K/2.5K, 2.5K/0.5K, 2.5K/2.5K;
      saturated; 512 requests
    & Co-located: DP:PP:TP$=$1:1:8, 2:1:4, 4:1:2;
      disaggregated: 1:1:4@1:1:4, 2:1:2@2:1:2; max batch$=$64
    \\

    \S\ref{sec:stage_ablation} Stage-wise
    & Train: Qwen2.5-72B-Instruct, Qwen3-32B; test: Qwen3-14B
    & vLLM
    & H200
    & ShareGPT; ISL/OSL$=$2.5K/0.5K, 0.5K/2.5K, 15K/0.1K;
      saturated; 64 requests
    & Train TP$=$2; test TP$=$1; max batch$=\{8,32\}$
    \\

    \S\ref{sec:control_fidelity} Native control
    & Qwen3-32B
    & vLLM
    & H200
    & ShareGPT; ISL/OSL$=$16.3K/1K; 100 req/s; 16 requests
    & DP:PP:TP$=$1:1:1; max batch$=$16;
      modified max-KV / original FCFS preemption
    \\

    \S\ref{sec:traffic_ablation} Closed-loop traffic
    & Qwen3-32B
    & vLLM
    & H200
    & ShareGPT; ISL/OSL$=$0.5K/2.5K;
      rate$=$10 req/s, saturated; 64 requests
    & DP:PP:TP$=$1:1:1; max client concurrency$=$8
    \\

    \S\ref{sec:coalescing_ablation} Coalescing
    & Qwen2.5-72B-Instruct
    & vLLM
    & H200 profile
    & ShareGPT; ISL/OSL$=$2.5K/0.5K, 2.5K/2.5K;
      saturated; 512 requests
    & DP:PP:TP$=$1:1:8; max batch$=$256;
      coalescing on/off
    \\

    \S\ref{sec:prefill} Long-context
    & Qwen3-235B-A22B
    & SGLang
    & B200
    & GSP~\cite{sglang_bench_serving}; ISL$=$2.5K--20K sweep; OSL$=$50;
      1 req/s; 128 requests; prefix hit$=\{0,20,40,60,80\}\%$
    & DP:PP:TP$=$1:1:8; max batch$=$16
    \\

    \S\ref{sec:generalization} Generalization
    & Llama-3.1-70B\textsuperscript{D};
      QwQ-32B and GPT-OSS-120B (held out)
    & vLLM / SGLang
    & B200
    & ShareGPT\textsuperscript{D} / RepoQA / AIME 2026;
      ISL/OSL$=$1--2.5K sweep
      (2.5K/2.5K\textsuperscript{D});
      rate$=$\{5,10,50,100\} req/s, saturated\textsuperscript{D};
      512 requests
    & DP:PP:TP$=$1:1:8; max-batch sweep (vLLM: 256\textsuperscript{D}; SGLang: 64\textsuperscript{D})
    \\

    \S\ref{sec:scalability} Scaling
    & Llama-3.1-70B
    & vLLM
    & B200
    & ShareGPT; ISL/OSL$=$2.5K/2.5K;
      saturated; 512 requests
    & DP:PP:TP$=$1:1:8; max batch$=$256;
      CPU cores$=$2--32
    \\

    \S\ref{sec:use_case} Search
    & Qwen3-32B
    & SGLang
    & B200
    & ShareGPT; ISL/OSL$=$2.5K/2.5K; saturated; 512 requests
    & Co-located / disaggregated; 228 candidate configurations; max batch=64
    \\

    \S\ref{sec:patching} Diagnosis
    & Qwen3-32B
    & SGLang
    & B200
    & ShareGPT; ISL/OSL$=$2.5K/2.5K; saturated; 512 requests
    & DP:PP:TP$=$1:1:8; max batch$=$32--512
    \\

    \bottomrule
  \end{tabularx}
\end{table*}

\end{document}